\documentclass[11pt,a4paper]{article}
\pdfoutput=1
\usepackage[usenames,dvipsnames]{xcolor}
\usepackage{jheppubx}
\usepackage{enumitem}
\usepackage{graphicx}
\usepackage{dcolumn}
\usepackage{bm,psfrag}
\usepackage{overpic}

\usepackage[utf8]{inputenc}

\usepackage{amsmath}
\usepackage{cleveref}
\usepackage{subfigure}
\usepackage{float}
\usepackage[sf,isu]{caption}
\usepackage{bbm}

\usepackage[mathscr]{euscript}

\usepackage{setspace}

\hypersetup{pdftitle={Monstrous entanglement},    
	pdfauthor={ddp}, 
	colorlinks=true, 
	linkcolor=blue!70!black, 
	citecolor=blue!39!black!,
	filecolor=OliveGreen, 
	urlcolor=RoyalBlue!60!black}

 \usepackage{overpic}
\def\bra#1{{\langle}#1|}
\def\ket#1{|#1\rangle}
\def\Tr{{{\rm Tr}}}
\def\cX{{\cal X}}
\newcommand{\be}{\begin{equation}}
\newcommand{\ee}{\end{equation}}

\usepackage{chngcntr}

\makeatletter
\g@addto@macro\bfseries{\boldmath}
\makeatother

\newcommand{\bea}{\begin{eqnarray}}
\newcommand{\eea}{\end{eqnarray}}
\newcommand{\ba}{\begin{eqnarray}}
\newcommand{\ea}{\end{eqnarray}}
\newcommand{\nn}{\nonumber \\}

\newcommand{\beq}{\begin{equation}}
\newcommand{\eeq}{\end{equation}}
\newcommand{\beqa}{\begin{eqnarray}}
\newcommand{\eeqa}{\end{eqnarray}}
\newcommand{\beqar}{\begin{eqnarray*}}
\newcommand{\eeqar}{\end{eqnarray*}}

\newcommand{\eg}{{\it e.g.,}\ }
\newcommand{\ie}{{\it i.e.,}\ }

\def\ausricht{\begin{aligned}}
\def\endeausricht{\end{aligned}}

\def\bZ {\mathbb{Z}}

\def\bM {\mathbb{M}}

\def\tr{\rm tr}

\def\t6 {T_\mt{D6}}

\newcommand{\mt}[1]{\textrm{\tiny #1}}

\newcommand{\cA}{{\cal A}}

\def\cale         {{\cal E}}

\def\ee           {{\rm e}}

\def\tr           {\mathop{\rm Tr}}

\def\Im           {{\rm Im\hskip0.1em}}

\def\sqr#1#2{{\vcenter{\vbox{\hrule height.#2pt
 \hbox{\vrule width.#2pt height#1pt \kern#1pt
 \vrule width.#2pt}\hrule height.#2pt}}}}

\usepackage{lipsum} 

\def\ee{\cale}

\def\aa1{\phi}
\def\cc1{\psi}

\def\k{\kappa}

\def\vev#1{\langle #1 \rangle}

\def\k{\kappa}

\def\vev#1{\langle{#1}\rangle}

\def\pd{\partial}
\def\cA{\mathcal{A}}
\def\cB{\mathcal{B}}
\def\cD{\mathcal{D}}

\def\makeitsmall{\begin{footnotesize}}
\def\endmakeitsmall{\end{footnotesize}}

\def\nn{\nonumber}
\def\Renyi{\text{R\'{e}nyi}~}
\def\Renyis{\text{R\'{e}nyis}}
\def\mm{V^\natural}

\newcommand{\normord}[1]{\vcentcolon\mathrel{#1}\vcentcolon}
\providecommand{\vcentcolon}{\mathrel{\mathop{:}}}

\def\blue#1{\textcolor{blue!90!black}{#1}}

\def\blue#1{\textcolor{black}{#1}}
\begin{document}


\title{Monstrous entanglement}

\newcommand{\batman}{\includegraphics[trim = 0mm 0mm 0mm 0mm, clip,scale=0.015]{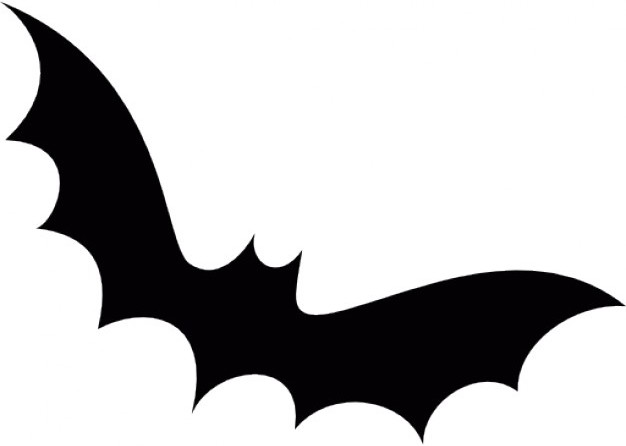}}

\author[\batman]{Diptarka Das}
\author[\reflectbox{\batman}]{\!, Shouvik Datta}
\author[\batman]{and Sridip Pal}

 \vspace{.1cm}
 \affiliation[\batman]{\vspace{.7cm}Department of Physics,\\ University of California San Diego,\\
 	9500 Gilman Drive, 
 	La Jolla, CA 92093, USA.\\ \vspace{-.2cm}
 }
 
 \affiliation[\reflectbox{\batman}]{Institut f\"{u}r Theoretische Physik,\\ Eidgen\"{o}ssische Technische Hochschule Z\"{u}rich, \\ 
 	Wolfgang-Pauli-Strasse 27,	8049 Z\"{u}rich, Switzerland.\\  \vspace{-0.2cm}
 }
%
%
%

\emailAdd{didas@ucsd.edu}
\emailAdd{shdatta@ethz.ch}
\emailAdd{srpal@ucsd.edu}

\abstract{The Monster CFT plays an important role in moonshine and is also conjectured to be the holographic dual to pure gravity in AdS$_3$.
We investigate the entanglement and R\'{e}nyi entropies of this theory along with other extremal CFTs. The \Renyi  entropies of a single interval on the torus are evaluated using the short interval expansion. 
 Each order in the expansion contains closed form expressions in the modular parameter. The leading terms in the $q$-series are shown to precisely agree with the universal corrections to  R\'{e}nyi entropies at low temperatures. Furthermore, these results are shown to match with bulk computations of \Renyi entropy using the one-loop partition function {on} handlebodies. 
 We also explore some features of \Renyi entropies of two intervals on the plane.
 }

{
\hypersetup{urlcolor= RoyalBlue!30!black}
\setlength{\parskip}{2pt}
\maketitle
\hypersetup{urlcolor=RoyalBlue!60!black}
}
\section{Introduction}

%
Monstrous moonshine was born from a mysterious observation, and it eventually grew into a  set of profound connections between modular forms, sporadic groups and conformal field theory \cite{conway-norton,FLM,Dixon-Ginsparg-Harvey,borcherds1992monstrous,gannon2006moonshine,duncan2015moonshine,Kachru:2016nty}. One of the crucial players in the story is that of the moonshine module $\mm$, which has the Fisher-Griess monster, $\mathbb{M}$, as its automorphism group. 
The moonshine module can be realized as the CFT of 24 free bosons compactified on the self-dual Leech lattice, along with a $\mathbb{Z}_2$ orbifolding \cite{FLM,Dixon-Ginsparg-Harvey}. 
Apart from  playing a key role in proving the Monstrous Moonshine conjecture, this CFT (along with its `extremal' generalizations) has also been proposed to be a candidate dual for pure gravity in AdS$_3$ \cite{Witten3d} (or its chiral version \cite{Li:2008dq}).   In this work, we begin an exploration into the information theoretic properties of this  theory. 

Entanglement entropy captures how quantum information is distributed across the Hilbert space of the system \cite{Amico:2007ag,Calabrese:2009qy,Ryu2006a}. Fundamentally, since the Hilbert space of the Monster CFT has an enormous group of symmetries (\ie of the largest sporadic group), one might expect manifestations of this in studies of entanglement entropy.
The huge symmetry, $\mathbb{M}$, is also expected to have an incarnation in the microstates of the conjectured bulk dual \cite{Witten3d}. This is something which has not been understood fully yet; see although for developments along these lines \cite{Duncan:2009sq,duncan2015moonshine}. In recent years, entanglement entropy has served as a powerful tool in reconstructing the bulk geometry and equations of motion of gravity \cite{Swingle:2012wq, Qi:2013caa,Bianchi:2012v,Bhattacharya:2013bna,Faulkner:2014jva,Faulkner:2017tkh}. It is therefore natural to explore this quantity in this context. Moreover, ambiguities still exist regarding the existence of extremal CFTs for $k\geq 2$ \cite{Gaberdiel:2007ve,Gaiotto:2008jt,Bae:2016yna}. One can hope that studies of entanglement and its information theoretic inequalities (arising from unitarity, causality and the like) may shed further light on this issue by filtering out the allowed theories. 

In this paper, we study the entanglement and \Renyi entropies of extremal CFTs. In the first setup, the CFTs shall be put on the torus (of spatial periodicity, $L$, and temporal/thermal periodicity, $\beta$; the modular parameter is $\tau\equiv 2\pi i\beta/L$). This will allow us to extract finite-size corrections from the $q$-series\footnote{$q$ is the nome, which is related to the modular parameter by $q=e^{2\pi i \tau}$. In more physical terms, these encode the corrections owing to the finiteness of the both the spatial and temporal directions. The $q$-series is equivalent to a low/high temperature expansion.}; see \cite{Azeyanagi:2007bj,Herzog:2013py,Datta:2013hba,Chen:disc,Datta:2014zpa,Chen:2015uia,Chen:2015kua,Chen:2015cna,Lokhande:2015zma,Schnitzer:2015ira,Schnitzer:2016xaj,Mukhi:2017rex} for previous works. Moreover, this unravels the pieces  special to CFTs of this kind. The entropies shall be evaluated in the short interval expansion (SIE), which requires information of one-point functions on the torus \cite{Calabrese:2009ez,Calabrese:2010he,Chen:2016lbu}. As we shall demonstrate, the SIE is fully determined to some order by expectation values of quasi-primaries within the vacuum Virasoro module alone. These expectation values can, in turn, be fully determined by Ward identities on the torus. In other words, the entanglement and \Renyi entropies for a single interval on the torus, at the first few orders in interval length, turns out to be  completely determined by the spectrum. For instance, to the order we have calculated, the entanglement entropy of $\mm$ in the short interval expansion reads
\begin{align*}
S_E(\ell) \ = \ 4 \log {\ell \over \epsilon} \ + \  \sum_{m=1}^{3}u_{2m} \left({E_6(\tau) \over E_4(\tau) }{j(\tau)  \over j(\tau)-744}\right)^{m}\left(\ell\over L\right)^{2m} \ + \ {\mathrm O}(\ell^8/L^8) . 
\end{align*}
Here, $\ell$ is the interval length and $\epsilon$ is the ultraviolet cutoff.  $E_{6,4}$ are the Eisenstein series of weights 6 and 4 and $j(\tau)$ is the Klein $j$-invariant. $u_{2m}$ are some constant coefficients. An important check of our result constitutes of a precise agreement with the universal thermal corrections to the \Renyi entropy, derived in \cite{Cardy:2014jwa}. This is essentially  the next-to-leading term in the $q$-series. Note that, the regimes of investigation of our present work and that of \cite{Cardy:2014jwa} are somewhat different from each other   which makes the agreement rather non-trivial. The expressions in \cite{Cardy:2014jwa} are perturbative in $q$ and non-perturbative in the length of the entangling interval. On the other hand, our expressions are perturbative in the interval length and non-perturbative in $q$. Nevertheless, one can expand these results both in the interval length and the nome (which is the `region of overlap' of these two analyses) and verify consistency of the results. We shall also see  that the \Renyi entropies are sensitive to the details of the operator content of the theory, \ie it realizes the McKay decompositions into Monster irreps.


To investigate the SIE at higher orders, the calculations will involve the one-point functions of primaries on the torus. A necessary ingredient for this would be both the spectrum and the information of three point OPE coefficients of the Monster CFT. This is an interesting line of investigation which is sensitive to finer details about the theory. Since the torus 1-point functions of primaries transform as cusp modular forms, one may hope that, to some controllable order, these one-point functions may be fixed using modular properties alone. We have not attempted to do so here and hope to address it in  the near future.

Yet another striking confirmation of our results comes from holographic computations of \Renyi entropy using the techniques of \cite{Barrella:2013wja,Faulkner2013}.  The $AdS_3$ dual of the CFT replica manifold is given by handlebody solutions. These are quotients of $AdS_3$ by the Schottky group. Upon evaluating the tree level and one-loop contributions to the gravitational path integral on these geometries, we have been able to show remarkable agreement with our CFT results for \Renyi entropy (to the order which we have calculated in the SIE). Not only does this concurrence constitute as a powerful confirmation of our CFT calculations, but also serves as a novel verification substantiating the holographic conjecture of \cite{Witten3d}\footnote{A modification to the conjecture of \cite{Witten3d} has been proposed in \cite{Benjamin:2016aww}. The aspects of entanglement and \Renyi entropies, which we study here, are robust in the sense that they hold true with or without of the modifications.}.  Although this agreement is perturbative, it does require details of partition functions of arbitrary genus both on the CFT and gravity sides.

We also analyse the \Renyi entropies of  two intervals for these CFTs on a plane. In particular, we shall consider the second   \Renyi entropies. The resulting genus of the replica manifold, via the Riemann-Hurwitz formula, is 1.
Since the torus partition functions are exactly known for these theories, this information can be directly used to find the second \Renyis. One can also extract the mutual \Renyi information. We shall uncover some interesting features of the crossing symmetric point, \ie when the cross ratio is $x=1/2$ \cite{Hartman1,Headrick:2010zt}. The mapping of the replica surface to the torus and the knowledge of the extremal partition functions facilitates a non-perturbative (in the cross-ratio) expression for this quantity, which allows for an explicit verification of unitarity constraints, \cite{Casini:2010nn}. For the third \Renyi entropy, the genus 2 partition function is necessary. This has been evaluated in a series of works \cite{Tuite:1999id,Mason:2006dk,gaiotto-yin,Avramis:2007gx,Gaberdiel:2009rd,Gaberdiel:2010jf} and has been a subject of renewed interest \cite{Headrick:2015gba,Belin:2017nze,Keller:2017iql,Cho:2017fzo,Cardy:2017qhl}. Although  we do not delve in this direction, this information can be used to obtain the third \Renyi entropy of two disjoint intervals. 

The outline of this paper is as follows. In Section \ref{sec:SIE} we review the short interval expansion for  calculating \Renyi entropy on the torus. We specialize to the case of the Monster CFT on the torus and provide  details of the calculations in Section \ref{sec:EEonTheTorus}. The analysis is then extended and generalized to other extremal CFTs in Section \ref{sec:otherExtremal}. We reproduce   the CFT results from the bulk dual in Section \ref{sec:bulk}. An attempt to determine the entanglement entropy in closed form is discussed in Section \ref{sec:conjecture}. The second \Renyi entropy of two intervals on the plane is studied in Section \ref{sec:2ndRenyi}.
Section \ref{sec:conclusions} contains our conclusions and avenues for future research. A number of technical details and extensions are relegated to the appendices and are referred appropriately from the main text.

\section{Short interval expansion for \Renyi entropy}\label{sec:SIE}
We are interested in the \Renyi entropy of a single interval (of length $\ell$) for a chiral CFT on a rectangular torus, $\mathbb{T}^2 \equiv \mathbb{S}^1_L \times \mathbb{S}^1_\beta$, \ie with spatial and temporal perodicities $L$ and $\beta$ respectively. The modular parameter, $\tau$, is therefore $i\beta/L$.  In this section, we briefly review the formalism to evaluate the \Renyi entropy as a perturbative expansion in the limit of short interval length. Much of this analysis for a generic CFT has been advanced by the work of \cite{Chen:2016lbu}.  Upon spatial bipartitioning, the $n$-th \Renyi entropy of a subsystem $\mathscr{A}$   is given by, 
\begin{equation}\label{renyi-def}
S_n = \frac{1}{1-n} \log \Tr_{\mathscr{A}} [\rho_{\mathscr{A}}]^n,
\end{equation}
where, $\rho_{\mathscr{A}}$ is the reduced density matrix of the region $\mathscr{A}$ defined by partially tracing out the complementary region in the full density matrix, $\Tr_{{\mathscr{A}}'}\rho$. 
The entanglement  entropy is   the $n\rightarrow 1$ limit of \eqref{renyi-def}, \ie the von Neumann entropy of the corresponding reduced density matrix. 
\begin{align}
S_E = - \Tr [\rho_\mathscr{A}\log \rho_\mathscr{A}].
\end{align}
In the path integral representation, \eqref{renyi-def} can be written as the partition function $Z_n$ corresponding to the `replica manifold', which is a $n$-sheeted Riemann surface, glued along the entangling interval. In the present context, the replica manifold is of genus $n$, \'ala the Riemann-Hurwitz theorem. The \Renyi entropy can be rewritten as 
\begin{align}\label{renyi-def2}
S_n = \frac{1}{1-n} \log \frac{Z_n}{(Z_1)^n}. 
\end{align}
 In most cases, $Z_n$ can be obtained as a correlation function of twist and anti-twist operators $\sigma, \tilde{\sigma}$, inserted at the endpoints of the entangling interval \cite{Calabrese:2009qy}.  The twist operators have conformal dimension $h_\sigma = h_{\tilde{\sigma}} =\frac{c(n^2-1)}{24n}$. The general expression for the OPE of the twist operators is  \cite{Calabrese:2009ez,Calabrese:2010he,Chen:2016lbu},
\begin{equation}\label{twistope}
\sigma(z ) \tilde{\sigma}(0) = \frac{ c_n}{z^{2 h_\sigma} } \sum_K d_K \sum_{r \geq 0} \frac{ a^r_K}{r!}z^{h_K +r} \partial^r \Phi_K(0).
\end{equation}
The summation over $K$  in the above equation is over all the quasi-primary operators in the replicated CFT and $c_n$ is the normalization of the twist operators. \blue{The coefficients $a_K^r$ ar defined as}
\begin{align}
a_K^r \equiv \frac{C^r_{h_K+r-1}}{C^r_{2h_K+r-1}}\ , \qquad \text{where, } C^y_x= \frac{\Gamma(x+1)}{\Gamma(y+1)\Gamma(x-y+1)}. \nn 
\end{align}
 The coefficient $d_K$ can be determined by the one-point function on the $n$-sheeted Riemann surface $\mathcal{R}_{n,1}$ as,
\begin{equation}\label{dk-def}
d_K = \frac{1}{\alpha_K L^{h_K} }\lim_{z\rightarrow \infty} z^{2h_K} \vev{ \Phi_K (z)}_{\mathcal{R}_{n,1} }.
\end{equation}
Here, $\alpha_K$ is the normalization of $\Phi_K$ on the plane. In the context of our calculation, the twist operator correlation function needs to be calculated by setting $z = \ell$, equation~\eqref{twistope}. When $\ell/L \ll 1$, we can use the OPE to find the \Renyi entropies as an expansion in $\ell/L$.  This is the essence of the short interval expansion (SIE). Alternatively,  from the path integral itself, one can get the SIE more directly by the pinching limit of the \textit{cutting-sewing} construction of higher genus Riemann surfaces, as used in \cite{Calabrese:2010he,Sen:1990bt}. 

We shall now provide a few details regarding the operators appearing in the identity Virasoro module\footnote{In addition  to these, the one-point functions of the primaries also contribute to the short-interval expansion. However, as we shall explain in Sec \ref{sec:Monster}, in context of the Monster CFT, owing to modular properties, they do not appear in the  first few orders (in fact, upto order $\ell^{24}$).}.  At the lowest level, \ie level 2, we have just the stress tensor. In the subsequent levels, we have quasi-primaries built from powers of the stress tensor and its derivatives. At level 4, we have
\begin{align}\label{level4}
	\mathcal{A} = \ \normord{TT} - \,  \frac{3}{10} \pd^2 T \ . 
\end{align}
There are two quasi-primaries at level 6.
\begin{align}\label{level6}
	\cB &= \ \normord{(\pd T)(\pd T)} - \,  \frac{4}{5} \normord{(\pd^2 T)T} -\ \frac{1}{42} \pd^4 T, \\
	\cD &= \ \normord{T(\normord{TT})} - \frac{9}{10} \normord{(\pd^2 T)T} - \frac{1}{28} \pd^4 T + \frac{93}{70c+29} \cB \ . 
\end{align}
The higher level quasi primaries become increasingly important for larger values of the ratio $\ell/L$. The corresponding $d_K$ factors \eqref{dk-def} have been calculated in \cite{Chen:2016lbu}. For example, $d_T ={(n^2-1)/(12 n^2)}$. Since the quasiprimary $\Phi_K$ appearing with $d_K$ can be on any of the replicated tori, the multiplicative counting factor can be coupled with $d_K$ by defining the coefficient $b_K$. For instance, in the case of the stress tensor we have $b_T = n d_T$. When there are $p$ stress tensors, one has, 
 $b_{TT\dots T} = \sum _{j_1 < j_2 \dots <j_p} d_{TT\dots T}^{j_1 j_2\dots j_p}$, where the replica index, $j$ runs from $0$ to $n-1$. Below we list some of the $b_K$ coefficients that we will require \cite{Chen:2016lbu}
\begin{align} 
&b_T =  {n(n^2 - 1)\over 12 n^2}. \qquad b_\cA= \frac{n \left(n^2-1\right)^2}{288 n^4}. \qquad b_\cD = \frac{n \left(n^2-1\right)^3}{10368 n^6}. \nn \\
&b_\cB = \frac{n  (-\left(n^2-1\right)^2 \left(2 (35 c+61) n^2-93\right) )}{10368 (70 c+29) n^6}.\nn \\
&b_{TT} = \frac{\left(n^2-1\right) \left(5 c (n+1) (n-1)^2+2 \left(n^2+11\right)\right)}{1440 c n^3}.\label{b-coeff} \\
&b_{T\cA} = \frac{\left(n^2-1\right)^2 \left(5 c (n+1) (n-1)^2+4 \left(n^2+11\right)\right)}{17280 c n^5}. \nn \\
&b_{TTT} = \tfrac{(n-2) \left(n^2-1\right) \left(35 c^2 (n+1)^2 (n-1)^3+42 c \left(n^2-1\right) \left(n^2+11\right)-16 (n+2) \left(n^2+47\right)\right)}{362880 c^2 n^5}. \nn 
\end{align}
Collecting the leading order terms in the SIE for the $n$-replicated torus partition function, we have the following,
\begin{align}\label{sie-01}
 {Z}_n = {\rm tr}_A \left[(\rho_A)^n \right]\ = \  \ & \frac{c_n L^{2(h_\sigma)}}{\ell^{2(h_\sigma)}} \bigg[ 1+ b_T \vev{T}\frac{\ell^2}{L^2} +(b_\cA \vev{\cA} +b_{TT}\vev{T}^2)\frac{\ell^4}{L^4} \nn   \\
& \qquad \qquad  + \left( b_\cB \vev{\cB} +  b_\cD \vev{\cD} + b_{T\cA} \vev{T} \vev{\cA}+b_{TTT}\vev{T}^3 \right)\frac{\ell^6}{L^6} + \cdots \nn  \\
& \qquad \qquad + \sum_{\phi}\frac{ \ell^{2h_\phi}}{L^{2h_\phi}} \left( b_{\phi\phi} \vev{\phi}^2 + \cdots  \right) \bigg],
\end{align}
where contributions from the primaries and their descendants are contained in the $\phi$ summation and the `$\dots$' come with higher powers of $\ell/L$.  Note from \eqref{b-coeff} that a considerable simplification occurs in the $n \to 1$ limit. In particular,  in the identity Virasoro module, only the coefficients of the form, $b_{T_i T_j \dots T_k}$ survive. As a result, the contribution of the identity Virasoro module to the entanglement entropy is completely determined by $\vev{T}$ on the torus and its higher powers. We shall return to this feature in Section \ref{sec:conjecture}.

\section{\Renyi entropies of the Monster CFT on the torus}\label{sec:EEonTheTorus}
\subsection{The Monster CFT}
The Monster CFT (the $k=1$ extremal CFT or the moonshine module $\mm$) has been explicitly constructed by Frenkel, Lepowsky and Meurman \cite{FLM}; see also \cite{Dixon-Ginsparg-Harvey}. The construction involves 24 free bosons compactified on 24-dimensional self-dual the Leech lattice, $\mathbb{R}^{24}/\mathrm{L}$. This is followed by an asymmetric  $\mathbb{Z}_2$ orbifolding, which removes the states at level 1. The partition function of the theory reads\footnote{An equivalent depiction of this is that of an $\mathbb{Z}_2$ asymmetric orbifold of the bosonic string compactified on the Leech lattice.}
\begin{align}\label{k1-partition-fn}
Z(\tau) \ &= \ \frac{\Theta_{\text{Leech}}}{2\eta^{24}} + \frac{1}{2\eta^{24}} \left[ (\vartheta_3\vartheta_4)^{12} +  (\vartheta_2\vartheta_3)^{12} - (\vartheta_2\vartheta_4)^{12}  \right]  \ =\ j(\tau) -744  \\
&=\  \frac{1}{q} + 196884 q + 21493760 q^2+864299970    q^3+20245856256 q^4 + \cdots \nn 
\end{align}
In the first equality, the first term is the contribution from the untwisted sector of the $\bZ_2$ orbifold whilst the other terms are from the twisted sector. 
As is well-known, the coefficients of the $q$-series display `moonshine', \ie they can be decomposed in terms of dimensions of irreducible representations of the Fisher-Griess monster, $\mathbb{M}$. More importantly, it has been shown that automorphism group of the vertex operator algebra of the moonshine module is indeed  $\mathbb{M}$ \cite{borcherds1992monstrous}. 

The 196884 states at level 2 have the following decomposition in terms of irreps of $\mathbb{M}$,
$
 {196884} = {196883} + {1 } 
$, which is the McKay's equation.  
This means it decomposes into a trivial representation and the smallest non-trivial representation of $\mathbb{M}$. In the language of the conformal field theory, these are dimension 2 operators which constitute the lightest states of the theory. The trivial representation corresponds to the stress tensor. In addition, there are 196883 primary fields of weight 2. This decomposition of the reducible representation shall play an important role in \S\ref{CHcompare}.  

Since we are dealing with a chiral CFT, there is a possible dependence on the conformal frame  in which the \Renyi entropies are being calculated owing to anomalies \cite{Iqbal:2014tta, Iqbal:2015vka}. This results in an extra term which is frame-dependent. In what follows, we shall however ignore these contributions since they are not specifically important for the theories we are dealing with. 

\subsection{Torus 1-point functions}\label{subsec:1point}\label{sec:Monster}

As discussed in Section \ref{sec:SIE}, the short interval expansion for the thermal \Renyi entropy requires the one-point functions of operators on the torus. Let us first focus on the identity Virasoro module. The quasi-primaries    contributing in the SIE, till level 6, are $T, \cA, \cB, \cD$, as given in equations~\eqref{level4}, \eqref{level6}. 
We shall now elaborate how to evaluate the one-point functions of these quasi-primaries. The one-point function of stress tensor can be found from the partition function itself:
\begin{align}\label{expec-stress-tensor}
\vev{T} &= 2\pi i \pd_\tau \log Z \ . 
\end{align}
Substituting the partition function \eqref{k1-partition-fn} and upon using the identity for the derivative of the $j$-invariant \eqref{j-magic},  we have
\begin{align}\label{st-k1}
\vev{T}   &= \frac{4 \pi ^2 E_6}{E_4} \left(\frac{ j }{ j -744}\right) . 
\end{align}
As expected, the modular weight of the one-point function of a quasi-primary equals its conformal weight. We shall remark on this further below. 

The expectation values of normal ordered products of the stress tensor can be found as follows. We can first use the Ward identity on the torus to find the correlation function of multiple stress-tensors. The Ward identity, derived in \cite{Eguchi:1986sb,Felder:1989vx}, is as follows 
\begin{align}
&\vev{T(v) T(v_1) T(v_2)\cdots T(v_s)} - \vev{T(v)}\vev{T(v_1) T(v_2)\cdots T(v_s) } \nn \\
=  & \ \bigg\lbrace 2\pi i \pd_\tau + \sum_{j=1}^s \left[ 2(\wp(v-v_j)+2\eta_1) + (\zeta(v-v_j) +2\eta_1 v_j) \pd_{v_j} \right]  \bigg\rbrace  \vev{T(v_1) T(v_2)\cdots T(v_s) } \nn \\
&\quad + \sum_{j=1}^s \frac{c}{12} \wp''(v-v_j) \vev{T(v_1) \cdots T(v_{j-1})T(v_{j+1})\cdots T(v_s) }   . 
\end{align}
Here, $\wp(v)$ and $\zeta(v)$ are the Weierstra{\ss}  elliptic functions. Once the correlators are evaluated using the above, we find the expectation values of normal ordered powers of the stress tensor by taking the coincident limit of these operators and subtracting out the OPE singularities. For instance,
\begin{align}
\vev{\normord{TT}(v)} \equiv \lim\limits_{u\to v } \left[ \vev{T(u)T(v)} - \text{singular terms} \right]. 
\end{align}
Performing this procedure, we end up with the following expressions for the expectation values of the quasi-primaries at level 4 and 6  \cite{Chen:2016lbu}
\begin{align}\label{expec:QP}
\vev{\cA} &= \frac{c g_2}{120} + \left(\vev{T}+4\eta_1 +2\pi i \pd_\tau\right) \vev{T}, \nn  \\
\vev{\cB} &= - \frac{9 c  g_3}{70} - \frac{9 c g_3 }{25} \vev{T}, \\
\vev{\cD} &= - \frac{3c(5c+22)g_3}{5(70c+29)} + \frac{c g_2\eta_1}{15} + \frac{c}{60}\pi i \pd_\tau g_2 + 2\pi \left[ 3(\vev{T} +4\eta_1 )i\pd_\tau  - 2\pi \pd_\tau^2 \right]\vev{T}  \nn \\
& \ \ \ \,  + \frac{(42c^2 -61c - 836)g_2}{24(70c+29)} \vev{T} + \left[ (\vev{T}+4\eta_1)(\vev{T}+8\eta_1) + 8\pi i \pd_\tau \eta_1 \right] \vev{T}. \nn
\end{align}
Here, $\eta_1$ and $g_{2,3}$ are the Eisenstein series with different normalizations (see also Appendix \ref{app:A} for further details). 
\begin{align}
\eta_1 = 4\pi^2 E_2, \qquad g_2 = \frac{4\pi^4}{3} E_4, \qquad g_3 = \frac{8\pi^6}{27}E_6 .
\end{align}
Upon substituting the form of the partition function \eqref{k1-partition-fn} for the Monster CFT  in the above formulae,  we have the following expressions.
\begin{align}\label{monster-1-point}
\begin{aligned}
\vev{\cA} &= {4\pi^4 \over 15} \left[\frac{31 E_4(j-24) }{(j-744)} + \frac{40 E_6^2 j }{E_4^2(j-744)}\right] ,\\ 
\vev{\cB} &= -{496\pi^6 \over 175} \frac{ E_6(j-240) }{(j-744)} ,\\
\vev{\cD} &= -{8\pi^6} E_6 \left[ {93 (1823 j-16) \over 15381 (j-744)} + {27344 E_6^2 j \over 15381 E_4^3 (j-744) } \right].  
\end{aligned}
\end{align}
In the above expressions, the derivatives of the $j$-invariant and the Eisenstein series were simplified by making repeated use of the Ramanujan identities \eqref{Ramanujan}. Note that the leading terms ($q^0$ term) in the above expressions can also be derived by finding the appropriate Schwarzian derivatives for each of the quasi-primaries under conformal transformation from the plane to the cylinder. 

It is easy to see that the 1-point functions of the quasi-primaries are modular functions with their respective conformal weights.  Under modular transformation, the expectation value of an operator $\mathbb{O}$, (primary or quasi-primary) of weight $h$, transforms as 
\begin{align}\label{mod-trans}
\vev{\mathbb{O}}_{a\tau +b \over c\tau +d} \ = \ (c\tau + d) ^h \vev{\mathbb{O}}_\tau.
\end{align}
Furthermore, the $q$-series for the 1-point functions of the quasi-primaries within the vacuum Virasoro module starts out at $q^0$. These are actually \textit{\blue{meromorphic} modular forms} of weight $h$. (For example, \eqref{st-k1}  has a pole
in the upper half-plane at the point where the partition
function $j- 744$ vanishes.) On the other hand, the unnormalized expectation value\footnote{Here, `unnormalized' refers to the feature that the the quantity is not divided by the torus partition function. This is denoted by the prime ($'$) in \eqref{p-vev}. We thank the anonymous referee for clarifying comments on this paragraph.} of a primary, $\Phi$, on the torus is given by 
\begin{align}\label{p-vev}
 \vev{\Phi}' &= \tr\left[\Phi \, q^{L_0 -\frac{c}{24}} \right]  = \sum_{i} \vev{i |\Phi|i} q^{h_i - \frac{c}{24}} .
\end{align}
In the second equality, the sum is over all operators of the theory. The leading contribution to this torus 1-point function arises from the lightest primary, $\chi$. This is 
\begin{align}
\vev{\Phi}' = \vev{\chi |\Phi|\chi} q^{h_\chi - \frac{c}{24}} + \text{(higher powers of $q$)}.
\end{align}
Recall that in the moonshine module, all primary operators are of integer conformal dimension (and $h\geq 2$). The above $q$-series, therefore, starts out with a positive power of $q$. This fact combined with the modular transformation property \eqref{mod-trans}, implies that the unnormalized torus 1-point function of a primary is a \textit{cusp  form} of weight $h$. It is a well known fact no cusp  forms of weight less than 12 exist (see Appendix \ref{app:A} for further details). As a consequence, all torus 1-point functions of primaries with conformal weight less than 12 is zero \cite{Yin:partition,gaiotto-yin}! Hence, from the short interval expansion \eqref{sie-01}, the primary fields of the Monster CFT contribute only at order $(\ell/L)^{24}$ and higher. The 3-point coefficients do not play a role, until that order, and the R\'enyi entropies depend only on the spectrum. This  feature, arising purely from modular properties, greatly simplifies the analysis of \Renyi entropy for the first few orders in the short interval expansion\footnote{{This statement can be refined even further by using the Monster symmetry. It is possible that primaries of weights even higher than $12$ are the ones which contribute. This can be pinned down by finding which structure constants appearing in \eqref{p-vev} are non-zero. We thank Matthias Gaberdiel for pointing this out.}}.


\def\cX{\mathscr{M}}

\subsection{\Renyi and entanglement entropy in the SIE}\label{subsec:sie-torus}

We now have the necessary ingredients to write the \Renyi entropy in the short interval expansion. Using \eqref{renyi-def2}, \eqref{sie-01} and the results of the one-point functions of the quasiprimaries from \eqref{monster-1-point}, the short interval expansion can be organized as follows. $Z(\tau)$ appearing below is the partition function of the moonshine module, $j(\tau)-744$. 
\begin{align}\label{renyi-sys}
S_n (\ell)= \frac{2(n+1)}{n} \log \frac{\ell}{\epsilon} + \sum_{\k=1}^\infty \, {\cX_{2\k}(\tau) \over Z(\tau)^\k}  \left(\ell \over L\right)^{2\k} .
\end{align}
The leading term is the universal area-law term. 
$\cX_{2\k}(\tau)$ are  { modular functions} of weight $2\k$.  As mentioned earlier, we have calculated this till the sixth order. Explicitly, they are given by
\begin{align}
\cX_2(\tau) &= -\frac{\pi ^2 }{3 } \frac{ (n+1)  }{n}   \frac{E_6  }{ E_4   }   \mathscr{P}_{2,1}  , \\
\cX_4(\tau) &= \frac{\pi ^4 }{1080}\frac{ (n+1)  }{n^3}  \left[  {E_6^2 \over E_4^2} \mathscr{P}_{4,1}  - { }   E_4 Z \mathscr{P}_{4,2}\right],\\
\cX_6(\tau) &= \frac{\pi ^6 }{60840}\frac{ (n+1)  }{n^5} \left[ {E_6^3 \over E_4^3} \mathscr{P}_{6,1}  - {  }   E_6  Z \mathscr{P}_{6,2}\right].
\end{align}
Notice that the structure of the above expressions are in terms of a basis of weight $2\kappa$  {`almost modular forms'} built of the Eisenstein series $E_4$ and $E_6$ \footnote{The Eisenstein series $E_{4,6}$ are holomorphic modular forms, whilst the ratio $E_6/E_4$ is an almost modular form. Since $E_4$ becomes $0$ at $\tau=\rho\equiv \exp\left(\tfrac{2\pi i}{3}\right)$, the holomorphicity breaks down at $\tau=\rho$. The {\it almost modular form} is, in fact, a generalization of modular forms and is polynomial in $({\Im[\tau]})^{-1}$ with coefficients being holomorphic function of $\tau$. This reduces to the standard modular form when the polynomial is of degree zero. 
	}. The functions $\mathscr{P}_{2\kappa,i}$ are modular functions, which are polynomials of the $j$-invariant. The coefficients of these polynomials depend on the \Renyi index $n$.  These have the forms
\begin{align}
\mathscr{P}_{2,1} &= j  ,\\
\mathscr{P}_{4,1} &=  \left(19 n^2-31\right)j^2+29760 \left(n^2-1\right)j ,\qquad \qquad \mathscr{P}_{4,2} = 31 (n^2-1)(j-24)  ,
 \nn \\
\mathscr{P}_{6,1} &= \left(79 n^4-341 n^2+310\right)j^2 +6944  \left(67 n^4-170 n^2+103\right)j+51663360 \left(n^2-1\right)^2 ,\nn  \\
\mathscr{P}_{6,2} &= 31  (n^2-1) \left( \left(n^2-10\right)j^2+8 \left(1753 n^2-1654\right)j -5952 \left(19 n^2-1\right)\right) .\nn 
\end{align}

The entanglement entropy is given by the $n\to 1$ limit of \eqref{renyi-sys}. It takes a rather simple form, since the 1-point functions of the stress tensor is the only contribution at the first few orders; equation \eqref{sie-01}. This is because the $b_{\mathcal{O}}$ coefficients -- equation \eqref{b-coeff} -- for the other quasiprimaries vanish in this limit. Explicitly, the entanglement entropy is 
\begin{align}\label{sieEE}
S_E (\ell)\ =\  4 &\log \frac{\ell}{\epsilon} \ -\ \frac{2 \pi ^2  }{3 }\frac{ E_6  }{ E_4 }\frac{  j }{ (j -744)} \frac{\ell^2}{L^2}\ -\  \frac{\pi^4}{45} \left[\frac{ E_6  }{ E_4 }\frac{  j }{ (j -744)}\right]^2 \frac{\ell^4}{L^4} \nn \\&
- \  \frac{4\pi^6}{2835} \left[\frac{ E_6  }{ E_4 }\frac{  j }{ (j -744)}\right]^3\frac{\ell^6}{L^6}\ +\ {\rm O}\left(\frac{\ell^6}{L^6}\right). 
\end{align}


As mentioned earlier, the SIE contains terms which are closed form expressions in the modular parameter of the torus (or, all orders in the $q$-expansion). In order to facilitate a comparison with the universal predictions of \cite{Cardy:2014jwa} and with holography (to be discussed in Section \ref{sec:bulk}), we require the $q$-expansion of \Renyi entropy \eqref{renyi-sys}. This is given as follows
\def\bS{\mathcal{S}}
\begin{align}\label{cft-renyi}
S_n = \bS_0 +\bS_2 q^2 + \bS_3 q^3 +\cdots ,
\end{align}
where
\begin{align}
\bS_0 &= \frac{2 (n+1)}{n} \Bigg[\log \left(\frac{\ell}{\epsilon}\right)-\frac{1}{3 }\left(\pi\ell \over L\right)^2-\frac{1}{90 }\left(\pi\ell \over L\right)^4-\frac{2}{2835 }\left(\pi\ell \over L\right)^6+{\rm O}(\ell^8/L^8)\Bigg] ,\label{universal}\\ 
\bS_2  &= \frac{n+1}{n}\Bigg[131256   \left(\pi\ell \over L\right)^2 -\frac{8  \left(73834 n^2-90241\right)}{15 n^2}\left(\pi\ell \over L\right)^4
\nn  \\
&\qquad\qquad +\frac{8  \left(278954 n^4-754757 n^2+508617\right)}{315 n^4}\left(\pi\ell \over L\right)^6 + {\rm O}(\ell^8/L^8)   \Bigg]   \label{matchstick} ,\\ 
\bS_3 &=\frac{n+1}{n} \Bigg[21493760 \left(\pi\ell \over L\right)^2-\frac{128\left(237139 n^2-270723\right)}{3 n^2}\left(\pi\ell \over L\right)^4  \nn \\
&\qquad \qquad+\frac{64  \left(44107429 n^4-107271470 n^2+65179081\right)}{945 n^4}\left(\pi\ell \over L\right)^6+ {\rm O}(\ell^8/L^8)   \Bigg] . \label{fingers}
\end{align}The expression for $\bS_0$ is none other than the short interval expansion of universal contribution to the \Renyi entropy (with $c=24$), which is $$S_{\text{univ}}=\frac{c (n+1)}{12n} \log \Bigg|{L\over \pi\epsilon }\sin \left(\frac{\pi \ell}{L}\right)\Bigg|.$$
wherein, we have written down just the chiral/holomorphic contribution. 

\subsection{Leading terms in the $q$-series of \Renyi entropy}\label{CHcompare}

It has been proved in \cite{Cardy:2014jwa} that the leading finite-size correction in the low temperature expansion for the \Renyi entropy on the torus is universal. The \Renyi entropy takes the following form \cite{Cardy:2014jwa} in the low temperature expansion (with $h'>h$ below)
\begin{align}\label{CHp}
&S_n = \frac{c(n+1)}{12n} \log \bigg| \frac{L}{\pi \epsilon}\sin \frac{\pi\ell}{L}   \bigg| +g \, \delta S_n^{(\psi)}+ {\rm O}(e^{-2\pi h' \beta /L }).
\end{align}
where, the universal thermal correction is from the lightest primary, $\psi$ of weight $h$. 
\begin{align}\label{s-p}
\delta S_n^{(\psi)} =  \frac{1}{1-n} \bigg[\frac{1}{n^{2h-1}} \frac{\sin^{2h} (\pi \ell /L)}{\sin^{2h} (\pi \ell /nL)} -n \bigg]e^{-2\pi h \beta /L} . 
\end{align}
 Here, $g$ is the degeneracy of the lightest primary operator in the theory. The above equation and the ones to follow below have been modified for the case of the chiral CFT. 

The above formula \eqref{s-p} is derived by considering the low lying spectrum of operators and their contribution to the reduced density matrix. The  $n$th  {moment} of the reduced density matrix is 
\begin{align}\label{rho-n}
\tr (\rho_{\mathscr{A}})^n = \frac{\tr \left[ \tr_{{\mathscr{A}}'}(\ket{0}\bra{0}+\sum_{\psi} \ket{\psi}\bra{\psi}e^{-2\pi \beta h_{\psi}/L} + \cdots)\right]^n}{(1+g e^{-2\pi h_\psi \beta/L}+\cdots)^n} .
\end{align}
The summation in the numerator is over the lightest operators (primaries and/or quasi-primaries) of the theory; $g$ in the denominator is the degeneracy of the lightest states.  It has been assumed in \cite{Cardy:2014jwa} that the lightest state is a primary. 
The first subleading term in the above expression is then equivalent to the following two point function of the primary operator (see \cite{Cardy:2014jwa} for more details)
\begin{align}
{\vev{\psi(\infty)\psi(-\infty)}_n \over \vev{\psi(\infty)\psi(-\infty)}_1 } \  = \ \frac{1}{n^{2h}} {\sin^{2h}(\tfrac{\pi \ell}{L}) \over \sin^{2h} (\tfrac{\pi \ell}{nL}) } \  .
\end{align}
Using this in \eqref{rho-n} and, in turn, in the formula for the \Renyi entropy, yields \eqref{CHp}. 

The above analysis requires an appropriate modification if the lightest state is the stress-tensor, which is a quasiprimary. This has been pointed out in \cite{Chen:disc}. 
In such a situation, we have an additional term (instead of equation (20) of \cite{Cardy:2014jwa}). 
\begin{align}
{\vev{T(\infty)T(-\infty)}_n \over \vev{T(\infty)T(-\infty)}_1 } \  = \ \frac{1}{n^4} {\sin^4(\tfrac{\pi \ell}{L}) \over \sin^4 (\tfrac{\pi \ell}{nL}) } \ + \  \frac{c}{18}\left(1-\frac{1}{n^2}\right)^2 \sin ^4 \left(\frac{\pi \ell }{L}\right) . 
\end{align}
The second term arises from the Schwarzian derivative (of the uniformization transformation) which is not considered in \cite{Cardy:2014jwa}. This term leads to additional contributions to the the \Renyi entropy\footnote{See \eg equation (2.25) of \cite{Chen:disc}.}. Therefore, if the lightest operator is (or the set of such operators includes) the stress tensor, its contribution to  first finite-size correction is given by 
\begin{align}\label{s-t}
\delta S_n ^{(T)} = - \frac{c(n+1)^2(n-1)}{18n^3} \sin^4 \left(\frac{\pi \ell }{L}\right)q^2 +  \frac{1}{1-n} \bigg[\frac{1}{n^{3}} \frac{\sin^{4} (\pi \ell /L)}{\sin^{4} (\pi \ell /nL)} -n \bigg]q^2.
\end{align}
The first term in the above equation was obtained in the context of holographic entanglement entropy in \cite{Barrella:2013wja}. 

As mentioned earlier, for the moonshine module, the McKay decomposition of 196884 lightest operators  is into 196883 primaries, $\psi_i$, and the stress tensor. The leading finite-size correction therefore splits as 
\begin{align}\label{mckay-decomp}
\delta S_n \ = \ \delta S_n ^{(T)} \ +\  \sum_{i=1}^{196883}\delta S_n ^{(\psi_i)}  
\end{align}
Substituting \eqref{s-p} and \eqref{s-t} with $c=24$, we obtain 
\begin{align}\label{CHm}
\delta S_n = - \frac{4(n+1)^2(n-1)}{3n^3} \sin^4 \left(\frac{\pi \ell}{L}\right) q^2 +  \frac{196884}{1-n} \bigg[\frac{1}{n^{3}} \frac{\sin^{4 } (\pi \ell /L)}{\sin^{4} (\pi \ell /nL)} -n \bigg]q^2.
\end{align}
Note that the above result is non-perturbative in the interval length  in $\ell$ at each order in $q$. As we have alluded to in the introduction, this is somewhat complementary to the SIE we have derived in the previous sub-section. Nevertheless, we can expand the above result around $\ell/L\to 0$ and check it against the $q$-series of \eqref{cft-renyi}. 
The SIE at the order in $q^2$ from equation \eqref{CHm} reads as 
\begin{align}
\delta S_n \ =  &  \ \ (n+1) \Bigg[\frac{131256 }{n}  \left(\pi \ell \over L\right)^2       -\frac{8   \left(73834 n^2-90241\right)}{15 n^3}\left(\pi \ell \over L\right)^4
  \\
&\qquad \qquad +\frac{8   \left(278954 n^4-754757 n^2+508617\right)}{315 n^5}\left(\pi \ell \over L\right)^6 + {\rm O}(\ell^8/L^8)   \Bigg] q^2 . \nn 
\end{align}
This agrees precisely with the next-to-leading term in the $q$-expansion, as given in equation~\eqref{matchstick}. 

{It is possible to proceed further to higher orders in the $q$-expansion using this procedure\footnote{We thank Justin David for pointing this out.}; see \eg \cite{Chen:2015uia} which does it for the vacuum block. Since the low-lying spectrum of operators is explicitly known for the mooshine module, one can calculate the \Renyi entropies order by order in $q$ using the procedure of \cite{Chen:2015uia,Cardy:2014jwa}. For instance, at ${\rm O}(q^3)$ we need to consider the contribution from 21296876 primaries of weight 3, 196883 descendants  of the  primaries of weight 2 ($L_{-1} \psi$) and the derivative of the stress tensor $\pd T$ or $L_{-3}$ acting on the vacuum. This makes it clear that the McKay decomposition explicitly needs to be taken into account while calculating the \Renyi entropy (as in equation \eqref{mckay-decomp} for level 2). 
This feature, in some sense, makes moonshine more manifest in the $q$-series of the \Renyis. The operators at various levels contribute differently as functions of the interval length (depending on whether it is a primary, quasiprimary or a descendant), as opposed to the partition function (which treats the states appearing at each level on an equal footing). 
}

%

%

\subsection{Intermezzo : other vignettes}
\subsection*{McKay-Thompson series and twisted entanglement entropies}
In the spirit of the Monstrous moonshine conjecture \cite{conway-norton,borcherds1992monstrous}, we can also consider partition functions and their entanglement entropies corresponding to the McKay-Thompson series\footnote{The calculations of this subsection are simple generalizations of the preceding ones.}. These are essentially the partition functions twisted by an element $g \in \bM$ and  can be shown to be genus-0 Hauptmouduls. \blue{Physically, this twisting refers to the change in boundary conditions of generic fields of the theory along the temporal cycle : $ X(z+\tau)\mapsto g X(z)$}. Owing to cyclicity of the trace, these twisted partition functions are defined upto conjugation $h g h^{-1}$ and therefore there is one such quantity for each of the 194 conjugacy classes of $\bM$.   We consider the conjugacy class $2A$ as an example. The twisted partition function reads
\begin{align}
Z_{2A}(\tau) = \tr (g_{2A}q^{L_0 - \frac{c}{24}})=\frac{1}{q}  + 4372q + 96256q^2 +1240002q^3+10698752q^4+\cdots
\end{align}
Apart from having the genus-0 property, the coefficients also display moonshine for the Baby Monster, $\mathbb{B}$. The partition function actually admits a closed form expression involving $\vartheta$-functions
\begin{align}
Z_{2A}(\tau) = 16 \frac{\vartheta_3^2}{\vartheta_2^2} \left(\vartheta_3^4 + \vartheta_2^4 \over \vartheta_4^4\right)^4 -104. 
\end{align}
The analysis of \S\ref{subsec:sie-torus} can be performed analogously and  the {\it twisted} \Renyi entropy is found to be
\def\bAS{\mathcal{S}^{(2A)}}
\begin{align}
S_n^{(2A)} = \bAS_0 +\bAS_2 q^2 + \bAS_4 q^3 +\cdots 
\end{align}
where
\begin{align}
\bAS_0 &=\frac{2 (n+1)}{n} \Bigg[\log \left(\frac{\ell}{\epsilon}\right)-\frac{1}{3 }\left(\pi\ell \over L\right)^2-\frac{1}{90 }\left(\pi\ell \over L\right)^4-\frac{2}{2835 }\left(\pi\ell \over L\right)^6+{\rm O}(\ell^8/L^8)\Bigg] \label{t-universal}\\
\bAS_2  &=\frac{n+1}{n}\Bigg[\frac{8744}{3 } \left(\pi\ell \over L\right)^2-\frac{8  \left(4926 n^2-6019\right)}{45 n^2}\left(\pi\ell \over L\right)^4\nn
\\ &\qquad\qquad+\frac{8  \left(18686 n^4-50383 n^2+33883\right)}{945 n^4}\left(\pi\ell \over L\right)^6+ {\rm O}(\ell^8/L^8) \Bigg] \label{t-matchstick}\end{align}\begin{align}
\bAS_4 &=\frac{n+1}{n}\Bigg[96256 \left(\pi\ell \over L\right)^2-\frac{128 \left(16477 n^2-18733\right)}{45 n^2}\left(\pi\ell \over L\right)^4\nn \\
&\qquad\qquad+\frac{64  \left(217509 n^4-518702 n^2+310217\right)}{945 n^4}\left(\pi\ell \over L\right)^6+ {\rm O}(\ell^8/L^8) \Bigg]  \label{t-fingers}
\end{align}It can be checked that the above expression agree with predictions from universality of \cite{Cardy:2014jwa}. That is, \eqref{t-universal} and \eqref{t-matchstick} expressions can be reproduced by expanding the following about $\ell/L\to 0$. 
\begin{align}\label{t-CHm}
\delta S_n = - \frac{4(n+1)^2(n-1)}{3n^3} \sin^4 \left(\frac{\pi \ell}{L}\right) q^2 +  \frac{4372}{1-n} \bigg[\frac{1}{n^{2\Delta-1}} \frac{\sin^{2\Delta} (\pi \ell /L)}{\sin^{2\Delta} (\pi \ell /nL)} -n \bigg]q^2
\end{align}
Here, we have used the McKay decomposition for the operators at level 2, \ie there are 4371 primaries and the stress tensor at this level. 

\subsection*{Presence of spin-1 currents}
We can also consider meromorphic $c=24$ theories in which   spin-1 currents are present. The partition function is then of the form
\begin{align}
Z(\tau) = j(\tau) - 744 + N_1 \ .
\end{align}
It has been conjectured that there are 71 such theories with specific values of $N_1$ \cite{Schellekens:1992db}. The only modification which we need to make is $Z \mapsto Z+N_1$ in \S{\ref{subsec:sie-torus}}. It can also be checked to agree with the universal prediction for the thermal correction of \cite{Cardy:2014jwa}, which is now
\begin{align}
\delta S_n \ =\   \frac{N_1}{1-n} \bigg[\frac{1}{n} \frac{\sin^{2} (\pi \ell /L)}{\sin^{2} (\pi \ell /nL)} -n \bigg]\, q\, . 
\end{align}

\section{Extremal CFTs of $k\geq 2$}\label{sec:otherExtremal}
\subsection{Partition functions}
The form of the extremal partition function for arbitrary $k$  is given by the polynomial ($J(\tau)=j(\tau)-744$)
\begin{align}
%
{   {Z_k}(\tau) }=\ &J(\tau)^k -(196884 k-1) J(\tau)^{k-2} -(21493760 k-1) J(\tau)^{k-3}  \\&+\left(19381654728 k^2-59009461038 k+393770\right) J(\tau)^{k-4} \nn \\&+\left(4231777443840 k^2-16947377322260 k+43578174\right) J(\tau)^{k-5} + \cdots . \nn 
\end{align}
and this series stops at $J(\tau)^0$. The expansion above can be found by starting with a polynomial of $J(\tau)$ of degree $k$ and demanding the non-existence of primaries below the conformal dimension $k+1$.

There is another way to write the $k$th extremal partition function compactly using Hecke operators \cite{Witten:2007kt}
\begin{align}\label{hecke-equation}
Z_k(\tau)=\sum_{r=0}^k a_{-r} \mathsf{T}'_rJ(\tau).
\end{align}
Here, the coefficients $a_r$ are the degeneracies of the Virasoro vacuum character
\begin{align}
\chi_{\text{vac}}(\tau) \equiv \sum_{r=-k}^\infty a_r q^r = q^{-k} \prod_{n=2}^{\infty}\frac{1}{1-q^n}.
\end{align}
and the Hecke operator $\mathsf{T}'_r$ acting on a modular function $F(\tau)$ is defined as 
\begin{align}
\mathsf{T}'_s F(\tau) = \sum_{d|s}\sum_{b=0}^{d-1} F\left(\tfrac{s \tau+bd}{d^2}\right).
\end{align}
This implies that one may {expect} the partition function for higher $k$ also to have the symmetry of $\mathbb{M}$, since the Hecke-transformed $J$-function contains the dimensions of the irreps of $\mathbb{M}$ in its $q$-series \cite{Witten3d}. Nonetheless, it has been argued in \cite{Gaiotto:2008jt} that no $k=2$ extremal CFT may admit symmetry of $\mathbb{M}$.

Using the partition function above, we can calculate the short interval expansion of the \Renyi entropy as in Section \ref{sec:SIE}. It is worthwhile noting that the entanglement entropy up to the sixth order in the SIE is given by 
\begin{align}
S_E (\ell)\ =\  4k &\log \frac{\ell}{\epsilon} \ -\frac{1}{6} \vev{T} \frac{\ell^2}{L^2}\ -\frac{1}{720k}  \vev{T}^2 \frac{\ell^4}{L^4} \ - \frac{1}{45360k^2}  \ \vev{T}^3\frac{\ell^6}{L^6}\ +\ {\rm O}\left(\frac{\ell^8}{L^8}\right) \ .
\end{align}
We have performed the analysis for \Renyi entropies of the extremal CFTs at $c=48$ and $72$ in Appendix \ref{app:4872}. In the following, we just consider the leading terms in the $q$-expansion and investigate some universal features at $k\geq 2$. 


\subsection{Leading terms in the $q$-series of \Renyi entropy}
The $q$-series of the partition function, $Z_k$, for $k\geq2$ is given by 
\begin{align}
Z_{k}=q^{-k}\left[1+q^2+{\rm O}(q^3)\right], \qquad \log Z_k= -k\log(q)+{\rm O}(q^2).
\end{align}
Consequently, the expectation values of the quasi-primaries is given by (from equations \eqref{expec-stress-tensor} and \eqref{expec:QP})
\begin{align}
\ausricht
\vev{T}&=4 \pi ^2 k-8 \pi ^2 q^2+{\rm O}\left(q^3\right)\\
\vev{\cA}&=\frac{4}{15} \pi ^4 \left(60 k+11\right)\left[ k+20q^2+{\rm O}(q^3)\right]\\
\vev{\cB}&=\frac{16}{175} \pi ^6 \left[-31 k+42 (2880 k+1) q^2+{\rm O}(q^3)\right]\\
\vev{\cD}&=\frac{8 \pi ^6 (42 k+17) (48 k-1) (60 k+11) }{9 (1680 k+29)}\left[k+66 q^2+{\rm O}(q^3)\right]
\endeausricht
\end{align}
The corresponding \Renyi entropy is then
\begin{align}\label{qexp}
S_n = \frac{2k(n+1)}{n} \log \frac{\ell}{\epsilon} + \sum_{\k=1}^\infty \, {\cX_{2\k}(\tau) \over Z(\tau)^\k}  \left(\ell \over L\right)^{2\k} 
\end{align}
where ${\cX_{2\k}(\tau) \over Z(\tau)^\k}$ is given as follows upto $q^2$:
\begin{align}
\ausricht
{\cX_{2}(\tau) \over Z(\tau)}&=\frac{(n+1)\pi^2}{n}\left[-\frac{  k}{3 n}+\frac{2 q^2}{3}+{\rm O}(q^3)\right],\\
{\cX_{4}(\tau) \over Z(\tau)^2}&=\frac{(n+1)\pi^4}{n}\left[-\frac{ k  }{90  }-\frac{ \left[(60 k+9)n^2-(60k+11)\right]q^2}{45 n^2}+{\rm O}(q^3)\right],\\
{\cX_{6}(\tau) \over Z(\tau)^3}&=\frac{(n+1)\pi^6}{n}\left[-\frac{2   k  }{2835  }-\frac{2 \left[(420 k+17) n^4-2 (210 k+23) n^2+31\right] q^2 }{945 n^4}+{\rm O}(q^3)\right].
\endeausricht
\end{align}
The leading correction to \Renyi entropy in low temperature comes solely from the stress tensor and is given by
\begin{align}\label{s-t2}
\delta S_n=\delta S_n ^{(T)} = - \frac{4k(n+1)^2(n-1)}{3n^3} \sin^4 (\tfrac{\pi \ell}{L}) q^2 +  \frac{1}{1-n} \bigg[\frac{1}{n^{3}} \frac{\sin^{4} (\pi \ell /L)}{\sin^{4} (\pi \ell /nL)} -n \bigg]q^2
\end{align}
which, in the small $\ell/L$ limit, matches with the terms of order $q^2$ in equation \eqref{qexp}.

\section{\Renyi entropy from the gravity dual}\label{sec:bulk}

In this section, we shall evaluate the \Renyi entropy of the moonshine module from the gravity dual \cite{Witten3d,Li:2008dq} with the goal of making contact with the CFT calculations of the previous sections. Recall that one of the major grounds of the holographic conjecture of  \cite{Witten3d} is that the Virasoro vacuum character alone is not modular invariant, that one needs to include black hole states and construct a modular invariant partition function. Let us briefly review the stream of logic here.

The vacuum character also equals the one-loop partition function of the graviton in AdS$_3$ \cite{Maloney:2007ud}. For a CFT with central $c=24k$ -- which is related to the three dimensional Newton's constant by the Brown-Henneaux formula, $c=3/2G_N$ -- the holomorphic partition function with the one-loop graviton determinant included is 
\begin{align}\label{Witten-com}
Z_0(q) = q^{-k} \prod_{n=2}^{\infty} \frac{1}{1-q^n}
\end{align}
To account for the presence of black holes in the theory (which have $L_0$ eigenvalues $\geq 0$), this should be modified as 
\begin{align}
Z_k(q) = q^{-k} \prod_{n=2}^{\infty} \frac{1}{1-q^n} + {\rm O}(q)
\end{align}
The modular function of the above form exists and is unique, and is given in terms of the $j$-invariant. Using the definition $J=j-744$, the partition function for the theory with $c=24k$ is given by polynomials of $J$. The coefficients of the polynomial can in turn be determined by demanding that the partition function is of the form \eqref{Witten-com}. For the case of $k=1$, this is simply $Z(q)=J(q)$. 
 Further support in favour of this partition function has been provided by interpreting the polynomials in $J$ arising from a sum over geometries, which are modular images of AdS$_3$ -- referred to as the `Farey tail' \cite{Dijkgraaf:2000fq,Manschot:2007zb,Manschot:2007ha,Iizuka:2015jma,Honda:2015mel,Duncan:2009sq}\footnote{More concretely, our holographic calculations is relevant in the context of the chiral gravity conjecture  \cite{Li:2008dq}. It can be shown that the partition function of extremal CFTs can be obtained from chiral gravity as a regularized sum over Euclidean geometries \cite{MSS}. Yet another recent proof of this using localization techniques has also been provided in \cite{Iizuka:2015jma,Honda:2015mel}.}. 
 
We shall focus only on the $k=1$ theory, \ie the holographic dual to the moonshine module. The analysis will proceed in a very similar manner for the theories with $k\geq 1$. The (holomorphic) partition function can be separated into the classical or tree-level piece and the one-loop contribution\footnote{Unlike \cite{Barrella:2013wja}, which uses Schottky quotient of the BTZ black hole, we implicitly consider the regularized sum over geometries as our starting point. In other words, the `gravity partition function of handlebodies' is the weighted sum over geometries with genus $n$ boundary.}. 
\begin{align}\label{breakup}
Z_{k=1}^{\rm tree}(q) = q^{-1} \qquad Z_{k=1}^{\rm 1-loop}(q) = 1+ 196884q^2 +21493760q^3 + \cdots
\end{align}
The formalism for calculating  \Renyi entropy from the bulk has been expounded in \cite{Faulkner:2013yia} and extended to include one-loop contributions in \cite{Barrella:2013wja}. The procedure consists of finding the Schottky uniformization of the replica geometry in the boundary $\mathbb{C}/\Gamma$. The Schottky group $\Gamma$ is a discrete sub-group of PSL$(2,\mathbb{C})$. For a Riemann surface of genus $g$, it is generated by the loxodromic generators $L_i$, where $i=1,2,\cdots,g$. Once $\Gamma$ is found, we need to find the partition function of the holographic dual on AdS$_3/\Gamma$ which has $\mathbb{C}/\Gamma$ as its conformal boundary. The quotients of AdS$_3$ by the Schottky group $\Gamma$ are handlebody geometries.

\def\cS{\mathcal{S}}
\def\St{\mathcal{S}^{\rm tree}}
\def\So{\mathcal{S}^{\rm 1-loop}}
The classical (order $c$) contribution can be obtained from studying  monodromies of the torus differential equation and then integrating the accessory parameter. We refer the reader to \cite{Barrella:2013wja,Chen:disc} for further details. For the entangling interval $\mathscr{A}$ given by $[-y,y]$, the result is (in a  low temperature expansion with $q=e^{-2\pi \beta/L}$ and with $c=24$)
\begin{align}
S^{\rm tree}_n =& \ \frac{2(n+1)}{n} \log \sin \left(2\pi y \over \beta\right) + {\rm const.} - \frac{4}{3} \frac{(n+1)(n^2-1)}{n^3} \sin^4\left(2\pi y\over L\right)q^2 \nn \\
& \ - \frac{16}{3} \frac{(n+1)(n^2-1)}{n^3} \sin^4\left(2\pi y\over L\right)\cos^2\left(2\pi y\over L\right)q^3 + {\rm O}(q^4)
\end{align}
The first term is the well-known universal contribution to the \Renyi entropy. The second term is exactly the first term in \eqref{CHm}, which arises from the Schwarzian derivatives of the uniformization transformation as explained earlier. For the sake of eventual comparison with CFT results, we organize the finite size corrections as follows
\begin{align}
S^{\rm tree}_n =& \
\St_0 + \St_2 q^2 +\St_3 q^3 + \cdots
\end{align}
with 
\begin{align}
\St_0 &= \frac{{2}(n+1)}{n} \log \sin \left(2\pi y \over \beta\right) + {\rm const.} \label{shoes}\\
\St_2 &= - \frac{4}{3} \frac{(n+1)(n^2-1)}{n^3} \sin^4\left(2\pi y\over L\right)  \label{ships} \\
\St_3 &= - \frac{16}{3} \frac{(n+1)(n^2-1)}{n^3} \sin^4\left(2\pi y\over L\right)\cos^2\left(2\pi y\over L\right)  \label{sealing-wax}
\end{align}

The prescription to find the one-loop contributions to the quotients AdS$_3/\Gamma$ is as follows \cite{Barrella:2013wja,Krasnov:2000zq}.  We need to find the set of representatives  of primitive conjugacy classes $\gamma \in \mathcal{P}$ of the Schottky group, $\Gamma$. This is done by constructing non-repeated words from the loxodromic generators $L_i$ and their inverses upto conjugation in $\Gamma$. The final step consists of calculating the largest eigenvalues ($q_\gamma$) of the words for each primitive conjugacy class, $\mathcal{P}$; these are then substituted as the arguments for one-loop determinants. 
The free energy then reads
\begin{align}
\log Z^{\rm 1-loop}_n = \sum_{\gamma \in \mathcal{P}} \log Z^{\text{1-loop}} (q_\gamma)
\end{align} 
For the case at hand, the one-loop contribution is given in equation \eqref{breakup}. Therefore
\begin{align}\label{dong-formula}
\log Z^{\rm 1-loop}_n = \sum_{\gamma \in \mathcal{P}} {\rm Re}\big[ 196884 q_\gamma^2 + 21493760 q_\gamma^3  +   {\rm O}(q_\gamma^4) \big]
\end{align}
For the quadratic and cubic orders which appear in the above equation, the contribution from single letter words is sufficient. 
The largest eigenvalue $q_\gamma$ of the single-letter word is given by (we have retained terms up to $q^{1/2}$, which will suffice for our purposes)
{
	{\makeitsmall{
	\begin{align}\label{largest-eigenvalue}
	&q_\gamma^{-1/2} \approx q^{-1/2}\Bigg[ \frac{n\left(u_y^{-1/n}-u_y^{1/n}\right)}{ \left(u_y^{-1}-u_y\right)} + \frac{q \left({u_y^{-1}}-u_y\right)}{n \left(u_y^{-1/n}-u_y^{1/n}\right)}  \times  \\ & \left(\frac{u_y^{-1/n}-u_y^{1/n} }{4\left(u_y^{-1}-u_y\right)^2}\left(u_y^{1/n} \left[n(u_y+u_y^{-1})-(u_y-u_y^{-1})\right]^2-u_y^{-1/n} \left[n(u_y+u_y^{-1})+(u_y-u_y^{-1})\right]^2\right)-1\right)\Bigg] \nn 
	\end{align}}
	\endmakeitsmall}
}\begin{figure}[!t]\centering
\includegraphics[scale=0.35]{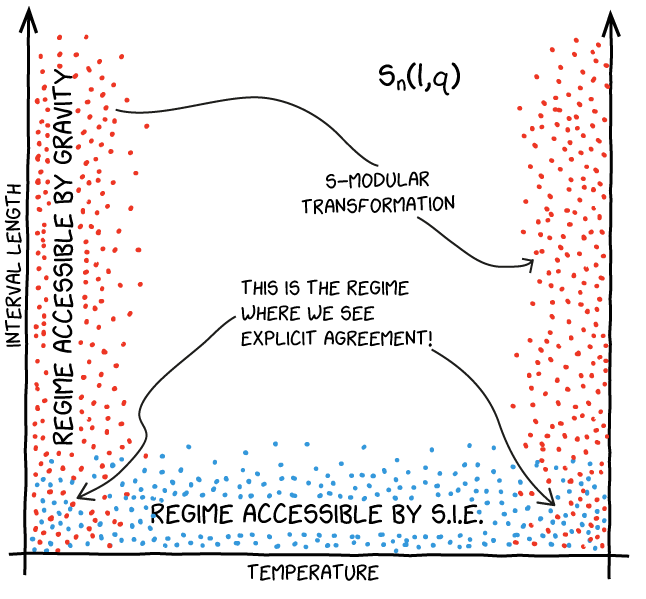}
\caption[blah]{Illustrating the accessible regimes of CFT and holographic calculations. The short interval expansion (SIE) is non-perturbative in $q$ (the nome) at each order in the interval length. On the other hand, the holographic calculation provides the $q$-expansion and is non-perturbative in the interval length at each order in $q$.  }
\label{fig:xkcd}
\end{figure} 
\hspace*{-.25cm}Here, $u_y=e^{2\pi y/L}$. These eigenvalues are independent of the index of the conjugacy class and therefore the sum in \eqref{dong-formula} is trivial. Substituting \eqref{largest-eigenvalue} in \eqref{dong-formula}, and using the path integral definition of the \Renyi entropy \eqref{renyi-def2}, we obtain
\begin{align}
S^{\rm 1-loop}_n = \So_2q^2 +\So_3q^3 + \cdots ,
\end{align}
where in terms of $\xi\equiv\tfrac{2\pi y}{L}$, we have
\begin{small}
	\begin{align}
	\So_2&=\frac{196884  }{(n-1) n^3} \left[n^4-{\sin ^4\left(\xi\right) \over \sin ^4\left(\frac{\xi}{n}\right)}\right], \label{cabbages}\\
	\So_3&=\frac{(n^2-1)^2}{(n-1)} \Bigg[ \frac{4 \sin ^4\left(\xi\right)}{n^5 \sin^6\left(\frac{\xi}{n}\right)}\left(-49221 \left(\frac{\sin \left(\frac{(n-1)\xi}{n}\right)}{n-1}- \frac{\sin \left(\frac{(n+1)\xi}{n }\right)}{n+1}\right)^2-\frac{5373440 \sin ^2\left(\xi\right)}{(n^2-1)^2}\right)  \nn \\ &\quad\quad\qquad\qquad+\frac{21493760 n}{(n^2-1)^2} \Bigg]. \label{kings}
	\end{align}
\end{small}In order to compare with the CFT expressions, we need to make the identification of the entangling interval, $\ell=2y$.
Combining the terms from the classical part \eqref{shoes}, \eqref{ships}, \eqref{sealing-wax} and those of 1-loop \eqref{cabbages}, \eqref{kings} it can been be seen that these precisely match with the short interval expansion calculated from the CFT. This is
\begin{align} 
\underset{\eqref{shoes}}{\St_0} \ &{\cong} \ \underset{\eqref{universal}}{\bS_0}\ \ ,\\
\underset{\eqref{ships}}{\St_2} + \underset{\eqref{cabbages}}{\So_2}\ &{\cong} \ \underset{\eqref{matchstick}}{\bS_2} \ \ , \\
\underset{\eqref{sealing-wax}}{\St_3} + \underset{\eqref{kings}}{\So_3}\  &{\cong}\  \underset{\eqref{fingers}}{\bS_3}\ \ . \label{finger-match} 
\end{align}
This agreement can be explicitly seen by Taylor expanding the LHS above about $\ell/L\to 0$. 
See also Fig.~\ref{fig:xkcd}
 which puts the CFT and holographic computations in a larger context and illustrates the regime where we are testing the bulk/boundary results\footnote{It is worthwhile to note that this agreement is not special for extremal CFTs. This is expected to be true in a more generic setting when the equivalence of CFT and AdS partition functions can be shown. We thank Justin David for pointing this out.}.
The first and second equalities above are well expected since they correspond to the universal contribution to the \Renyi entropy and is the universal thermal correction of \cite{Cardy:2014jwa} respectively. The third equality, equation \eqref{finger-match}, is substantially specific to the details of the CFT under consideration. A high temperature expansion for the \Renyi entropy can also be found upon the modular transformation 
$  
(L \mapsto i\beta, \ \beta \mapsto iL)
 $,
 which indeed agrees with the corresponding modular transformed $q$-series from the short interval expansion. 

 This holographic verification serves a two-fold purpose. Firstly, it confirms the correctness of our CFT expressions derived via the short-interval expansion. And secondly, it also provides a non-trivial verification of the holographic conjecture of \cite{Witten3d} and the one-loop exactness of the partition function in our context. Note that this agreement, albeit perturbative, is that of arbitrary genus $n$ partition functions of the CFT and the gravity theory\footnote{This was speculated as an optimistic possibility in \cite{MSS}.}. The moduli space in the present situation is, however, restricted to a plane -- consisting of $n$ replicas of a torus (each with modular parameter $\tau$) which are joined by tubes of the same pinching parameter (the interval length $\ell$). 

 We finally note that the low temperature expansion for the entanglement entropy (\ie the $n\to 1$ limit) as derived from holography is 
 \begin{align}\label{gravityEE}
 S_E =  \ 4 \log \left|\frac{L}{\pi\epsilon} \sin \left(\pi \ell \over L\right)\right| & \ +  4\times 196884 \left[1- {\pi \ell \over L}  \cot \left(\pi \ell \over L\right)\right]q^2  \\ &\ + 6\times 21493760 \left[1- {\pi \ell \over L}  \cot \left(\pi \ell \over L\right)\right]q^3 + {\rm O}(q^4).    \nn 
 \end{align}
where, we have fixed the `const.' in \eqref{shoes} by demanding that the universal term should be given by the Ryu-Takayanagi formula. 

\section{Towards a closed formula for entanglement entropy}\label{sec:conjecture}
The results for entanglement entropy which we have obtained from the short interval expansion, \eqref{sieEE}, and the $q$-series, equation \eqref{gravityEE}, obtained from holography agree in the $q\to 0$ and $\ell /L\to 0$ limit -- Fig.~\ref{fig:xkcd}. One may conjecture that these are expansions of the following expression 
\begin{align}\label{mauja}
\widetilde{S}_{E} \  \overset{\texttt{?}}{=} \ 4 \log \left| \frac{L}{\pi \epsilon} \left(\frac{E_6( \tau)}{E_4( \tau)}\frac{j( \tau)}{j( \tau)-744}\right)^{-1/2} \sin\left[\left(\frac{E_6( \tau)}{E_4( \tau)}\frac{j( \tau)}{j( \tau)-744}\right)^{1/2} \frac{\pi \ell}{L}\right] \right| . 
\end{align}
It can be checked explicitly that the $\ell/L$ expansion of this yields \eqref{sieEE} and, more remarkably,  the $q$-expansion  yields \eqref{gravityEE}, which is quite promising. The above formula is suited for a low temperature expansion, $\tau = i\beta /L$. It is clearly visible that the universal formula for entanglement entropy on a cylinder (with periodic spatial direction) \cite{Calabrese2009} can be recovered in the strict $q\to 0$ limit. 

The high temperature version of \eqref{mauja} can be obtained by a modular transformation. As we have noted earlier, 
\begin{align*}
\left(\frac{E_6( \tau)}{E_4( \tau)}\frac{j( \tau)}{j( \tau)-744}\right)  = \tau^{-2}  \left(\frac{E_6(-1/\tau)}{E_4(-1/\tau)}\frac{j(-1/\tau)}{j(-1/\tau)-744}\right). 
\end{align*}
Substituting this in \eqref{mauja} and recalling $\tau=i\beta/L$ and $j(-1/\tau)=j(\tau)$, we get
\begin{align}\label{mauja-high}
\widetilde{S}_{E} \ \overset{\texttt{?}}{=} \ 4 \log \left| \frac{\beta}{\pi \epsilon} \left(\frac{E_6(-\tfrac{1}{\tau})}{E_4(-\tfrac{1}{\tau})}\frac{j( {\tau})}{j( {\tau})-744}\right)^{-1/2} \sinh\left[\left(\frac{E_6(-\tfrac{1}{\tau})}{E_4(-\tfrac{1}{\tau})}\frac{j( {\tau})}{j({\tau})-744}\right)^{1/2} \frac{\pi \ell}{\beta}\right] \right| . 
\end{align}
which easily reproduces the universal formula at high temperatures in the cylinder limit; see \eg \cite{Calabrese2009}. 

Despite these niceties, it is rather unfortunate that the above formula \eqref{mauja-high} cannot be correct. It can be seen that it does not reproduce the thermal entropy in the limit $\ell\to L$, the limit when the sub-system equals the full system. This is the property \cite{Azeyanagi:2007bj}
\begin{align}\label{thermal}
S_E(\ell=L-\epsilon) - S(\ell=\epsilon) \ \overset{\epsilon\to 0}{=} \  S_{\text{thermal}} . 
\end{align}
The thermal entropy is, in turn, given by
\begin{align}
S_{\text{thermal}} = \log Z + \pd_\tau \log Z = \log(j(\tau)-744) - \left(\frac{E_6( \tau)}{E_4( \tau)}\frac{j( \tau)}{j( \tau)-744}\right). 
\end{align}
This disagreement with the proposed non-perturbative formula is plausibly related to the fact that the expressions -- \eqref{mauja} and \eqref{mauja-high} -- are not manifestly doubly periodic nor modular invariant.  In fact, one can explain why the conjectured formula is not entirely correct by looking at higher order terms in short interval (${\ell}/{L}$) expansion. We claim that the equation \eqref{mauja} captures the contribution to entanglement entropy coming from identity Virasoro module alone and does not capture contributions arising from conformal families of non-vacuum primaries.

{In $n\to 1$ limit, the short interval expansion \eqref{sie-01} yields the following expression for entanglement entropy 
\begin{align}\label{conj-exp}
S_E (\ell)\ =\  4 &\log \frac{\ell}{\epsilon} + \sum_{k} \tilde{b}_{T^k}\vev{T}^k\ \ell^{2k}  + \cdots,
\qquad \text{where, \ }
\tilde{b}_{T^{k}}= \underset{n\to1}{\text{lim}}  \, \frac{1}{n-1}\ b_{{\tiny{T^k}}} \ . 
\end{align}
the `$\cdots$' represent the contribution from the conformal families of non-vacuum primaries. As noted mentioned earlier, the torus 1-point function of all the primaries with conformal weight less than $12$ vanishes since cusp forms of lower modular weight do not exist. The extra terms denoted by `$\cdots$' in equation \eqref{conj-exp}, therefore, become relevant only at $ (\ell / L)^{24}$ and beyond. Hence,  the conjectured formula \eqref{mauja} is in fact a re-summation of identity Virasoro module, \ie 
\begin{align}\label{identity}
S_E \supset S_{\text{vac}} = 4 &\log \frac{\ell}{\epsilon}+\sum_{k} \tilde{b}_{T^k}\vev{T}^k\ \ell ^{2k}=\ 4 \log \left| \frac{2}{\epsilon\vev{T}^{1/2}}  \sin\left[\frac{\vev{T}^{1/2}}{2}\ell\right] \right| . 
\end{align} 
The proof of the above is simple in the cylinder limit of the torus (this is the zero-temperature limit which decompactifies the temporal direction, but keeps the periodicity of the spatial direction intact -- leading to $\mathbb{R}\times \mathbb{S}^1_L$). The one point function of primaries are $0$ on the cylinder. In other words, the entanglement entropy on the cylinder   receives contributions from the vacuum Virasoro module alone. 
\begin{align}\label{cyl1}
S^{\text{cylinder}}_E (\ell)\ =\  4 &\log \frac{\ell}{\epsilon} + \sum_{k} \tilde{b}_{T^k}\vev{T}^k_{\text{cylinder}}\ \ell^{2k} .
\end{align}
On the other hand, considering the two point function of twist operators and its conformal transformation to the cylinder, one can arrive at
\begin{align}\label{cyl2}
S^{\text{cylinder}}_E (\ell)\ &=\ 4 \log \left| \frac{L}{\pi \epsilon} \sin\left(\frac{\pi \ell}{L}\right) \right| =  4 \log \left| \frac{2}{\epsilon\vev{T}^{1/2}}  \sin\left[\frac{\vev{T}^{1/2}}{2}\ell\right] \right| . 
\end{align}
where, in the second equality we have used $\vev{T}_{\text{cylinder}} =   {4\pi^2}/{L^2}$. 
It can be seen that \eqref{cyl1} is precisely the Taylor expansion around $\ell/L\to 0$ of
 \eqref{cyl2}.  This leads to the cylinder-limit of the identity given in equation \eqref{identity} in which $S_{E}=S_{\mathrm{vac}}$, since one-point functions of non-vacuum primaries (and their descendants) vanish on the cylinder. Furthermore, one can obtain the high temperature version (with the `sinh' in \eqref{mauja-high}) by conformal transformation to the thermal cylinder, $\mathbb{R}\times \mathbb{S}^1_\beta$. 

The failure of the \eqref{thermal} can also be traced back to the fact that the conjectured formula is not true at $(\ell / L)^{24}$ and beyond, when cusp forms corresponding to 1-point functions of primaries contribute; as discussed in \S\ref{subsec:1point}. Therefore it does not reproduce the  expected behaviour when $\ell$ becomes of the same order as $L$.} Undoubtedly, a better approach which captures the entanglement entropy in all its entirety is desirable.

\section{\Renyi entropies of two intervals on a plane}\label{sec:2ndRenyi}
\subsection{Second \Renyi entropy and mutual information}
\begin{figure}[ht]
\hspace{-1.4cm}
 \includegraphics[scale=0.26]{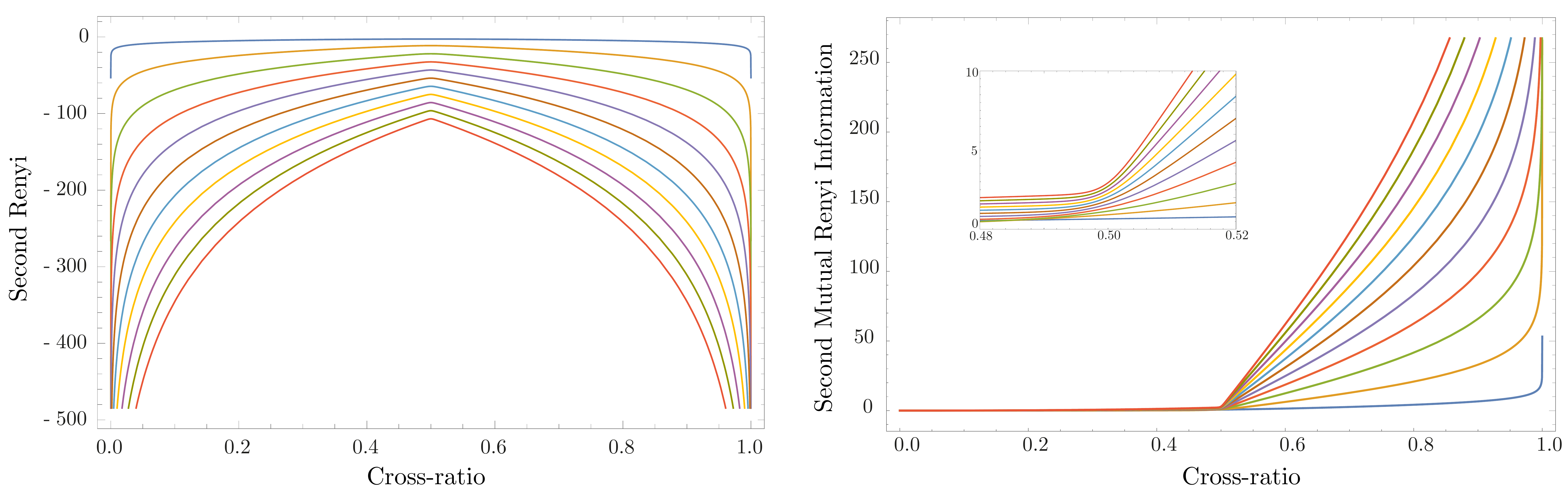}
\caption{ (Left)  Second \Renyi entropies and (Right) Mutual \Renyi Information for $k=1,5, 10, 15, 20, 25, 30, 35, 40, 45, 50, 55, 60$ as a function of $x$. The inset in (Right) shows the smooth behavior around $x=1/2$. }\label{figure-1}
\end{figure}

In this section, we aim to find out second \Renyi entropy and second mutual information of two disjoint subsystems given by the intervals $[u_1, v_1]$ and $[u_2,v_2]$. One can perform a conformal transformation
\begin{align}
w=\frac{(z-u_1)(v_2-u_2)}{(u_2-u_1)(v_2-z)}
\end{align}
so that the intervals $[u_1, v_1]$ and $[u_2,v_2]$ get mapped to $[0,x]$ and $[1,\infty]$, where $x$ is cross-ratio given by,
\begin{equation}
x = \frac{ ( v_1-u_1)(v_2 - u_2) } {(u_2-u_1)(v_2-v_1) }.\label{crossrat}
\end{equation}
%
The second \Renyi entropy for the intervals $[u_1, v_1]$ and $[u_2,v_2]$ can now be obtained as a function of $x$ using the four point function of  $\mathbb{Z}_2$ twist operators, $\sigma(z)$, having conformal weight $h_\sigma = c/16$. The appropriately normalized four point function of these twist operators, in turn, can be expressed in terms of the torus partition function of the CFT \cite{Headrick:2010zt}. For a chiral CFT, this is given by
\begin{eqnarray} \label{eqs2}
S_2 &=& - \log \Tr \rho^2 = - \log \vev{ \sigma(1) \sigma(0) \sigma(\infty) \sigma'(x) }, \nonumber \\
&=& - \log Z(\tau) + \frac{c}{24} \log \big[ 2^{8} x ( 1-x) \big]. \label{renyilunin}
\end{eqnarray}
Here the modular parameter $\tau$ is related to $x$ via,
\begin{equation}\label{taux}
\tau  = i  \frac{ {}_2 F_1 \big( \tfrac{1}{2}, \tfrac{1}{2} , 1, 1-x \big)}{{}_2 F_1 \big( \tfrac{1}{2}, \tfrac{1}{2} , 1, x \big)} .
\end{equation}
{The second term in \eqref{eqs2} takes into account the dependence on the conformal frame.}
Using the knowledge of the genus-1 partition functions for extremal CFTs, we  evaluate the second \Renyi entropy as a function of $x$, depicted in Fig. \ref{figure-1}a.
As the four point function of twist operators are crossing symmetric, we have $S_2(x)=S_2(1-x)$, implying that $S_2^{\prime}(1/2)=0$. This conforms to the Fig.~\ref{figure-1}a, where the maxima occurs at $x=1/2$ or equivalently at $\tau = i$. We also observe that the \Renyi entropy develops a sharper maxima with increasing $k$. From the dual gravitational perspective, the entanglement entropy undergoes a phase transition at $x=1/2$. However, it can be shown that $1/c$ corrections smoothen out the phase transition \cite{Barrella:2013wja}. The prominence of the peak therefore happens in the large $c$ regime when $1/c$ corrections are highly suppressed. Furthermore,  the absence of light primary states also contributes to such a {phase transition} at large $k$ (the spectrum of primaries is $h\geq k+1$)\footnote{This is related to fact that presence of too many light states wash out the Hawking-Page transition. This happens, for instance, in coset theories in $2d$ and vector models in 3$d$ \cite{Banerjee:2012aj}.}. The fact that the phase transition is indeed a large $c$ artifact can be seen from evaluating mutual information as well. The mutual information between the two subsystems $[u_1, v_1]$ and $[u_2,v_2]$, for a chiral CFT, is given by,
\begin{eqnarray}
I_2 &=& \log \big( x^{2h_\sigma} \vev{ \sigma(1) \sigma(0) \sigma(\infty) \sigma'(x) }\big) = \log Z(\tau) - \frac{c}{24} \log \bigg( 2^{8} \frac{ 1-x}{x^2} \bigg).
\end{eqnarray}
 
From the Fig.~\ref{figure-1}b, we observe that as $k$ increases, the second mutual \Renyi information, $I_2$, becomes flatter and goes to $0$ for $x\leq{1}/{2}$ followed by a sharp rise for $x\geq {1}/{2}$. The sharpness of the change (to be precise, the apparent discontinuity in the derivative of $I_2$ with respect to $x$ at $x={1}/{2}$) becomes more prominent as $k$ increases, corroborating to the large $c$ intuition. Nonetheless, since all the plots given are for finite $c$, there isn't any actual discontinuity in the derivative of $I_2$, as evident form the zoomed-in version of the graphs as shown inset of Fig.~\ref{figure-1}b. 

\subsubsection*{Second \Renyi maxima, $\left[S_2\right]_{\text{max}}$}  
\begin{figure}[!t]
 	\centering
	\includegraphics[scale=.7]{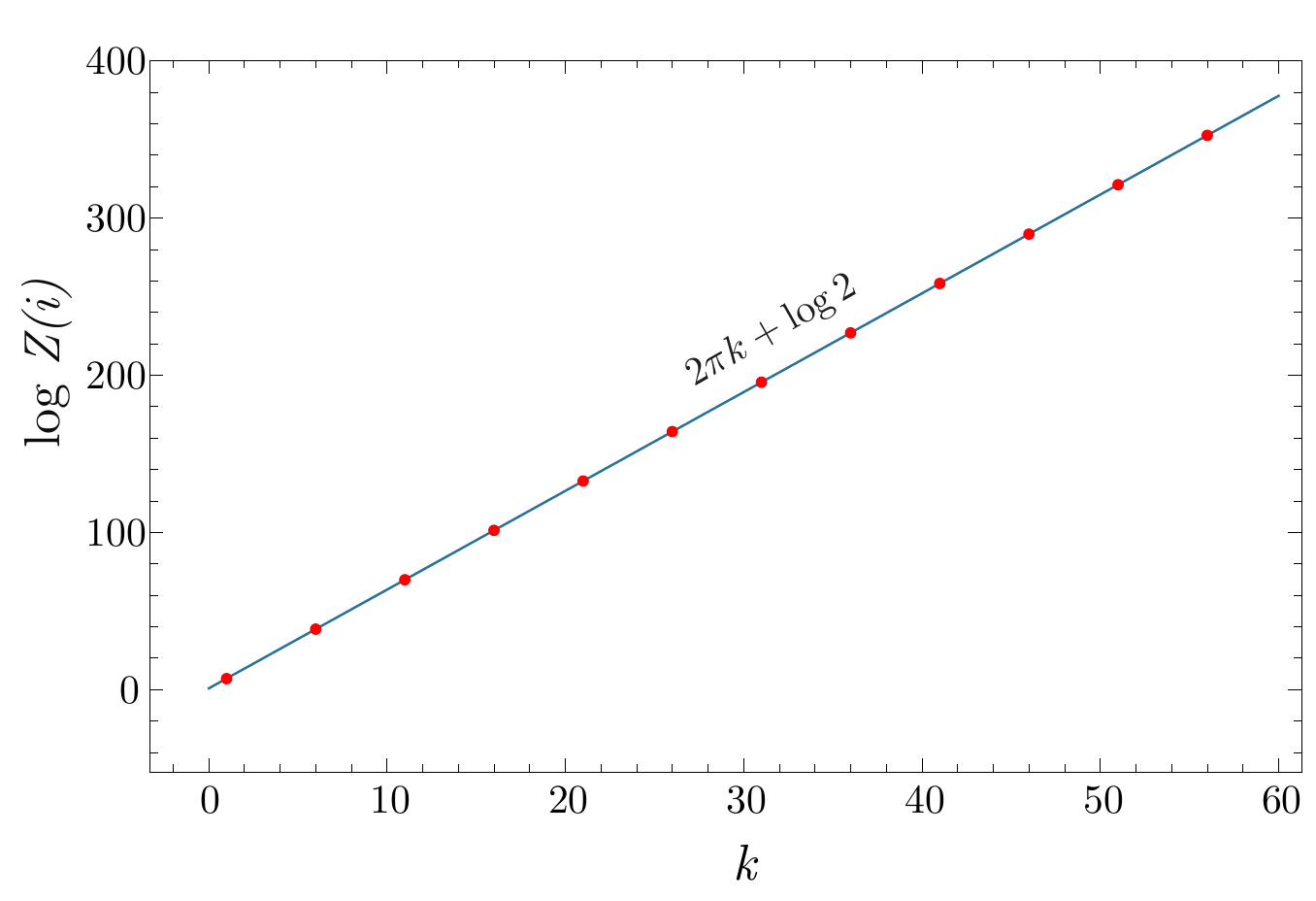}
\caption{Verifying $\log Z(i) \cong  { 2 \pi  k }   + \log 2 $ }\label{conj-test}
 \end{figure}

Curiously enough, the maximum value of \Renyi entropy of two intervals, as depicted in Fig.~\ref{figure-1}a, is approximately given by
\begin{align}
  \left[S_2 \right]_{\text{max}}  \ \cong \ 2k (3 \log 2- \pi ) - \log 2,
\end{align}
which arises from  
\begin{align} \label{estimate}
\log Z(i) \cong  { 2 \pi  k }   + \log 2 . 
\end{align}
We provide numerical evidence in favour of this in Fig.~\ref{conj-test}. 

At the self-dual point ($\tau=i$ at which $j(i)=1728$), we are therefore led to a rather non-trivial (but approximate) mathematical identity. Using the expression for the partition function of the $k$th extremal CFT in terms of the Hecke-operators acting on $j$, equation \eqref{hecke-equation}, we have 
\begin{align}\label{id-01}
\log\left[\sum_{r=0}^k a_{-r} \left[\mathsf{T}'_rJ(\tau)\right]_{\tau=i}\right] \ \ \cong \ \ 2\pi k + \log 2 . 
\end{align}
This approximation is true even at low values of $k$ and therefore isn't a statement about large central charge asymptotics. 
A crude derivation is as follows\footnote{We are grateful to Christoph Keller for this proof.}. {The partition function for the extremal CFT can be approximated as 
\begin{align}\label{zk}
Z_k(\tau) \approx q^{-k} + (\tilde{q})^{-k}
\end{align}
Here, $q=e^{2\pi i \tau}$ and $\tilde{q}=e^{-2\pi i /\tau}$. This is in the spirit of the `Farey tail' sum \cite{Dijkgraaf:2000fq}; although the partition function is invariant only under $S$-modular tranformations. From a holographic perspective, the above equation takes into account contributions from thermal AdS$_3$ and the BTZ black hole. Contributions arising from the other modular images of SL$(2,\mathbb{Z})$ are sub-dominant for purely imaginary $\tau$, especially when $k$ is large. At the self-dual point, $\tau=i$, the partition function \eqref{zk} is given by $Z_k(i) \approx 2\,e^{2\pi k}$, which is equivalent to equation \eqref{estimate}. 
}



 
\subsection{Constraints on second \Renyi}\label{7.2}
The existence of extremal CFTs beyond $k=1$ has a much debated status. One might hope to eliminate their possible existence (or perhaps save them from extinction!) by checking whether they satisfy all the inequalities involving \Renyi entropy. In this subsection, we take a small step in this direction using the inequalities derived in \cite{Casini:2010nn}\footnote{We thank Tarun Grover for discussions and for pointing out this reference sometime ago.}. 

The tool used in  \cite{Casini:2010nn} to constrain the \Renyi entropies is the wedge reflection positivity. This acts as follows, \blue{where $t$ and $y$ denote the Lorentzian coordinates within the wedge, \ie outside the light cone, see \cite[Fig.~1]{Casini:2010nn}}
\begin{align}(t,y)
\underset{\text{reflection}}{\underset{\text{Wedge}}{\longrightarrow}} (-t,-y) \ ,
\end{align}
and takes operator sub-algebra of one wedge ($y>0, |t|<y$) to its reflection ($y<0, |t|<-y$). In particular, it has been shown in \cite{Casini:2010nn} that $n$th \Renyi entropy satisfies the following inequality
\begin{align}\label{ineqC}
2S_n(A\bar B) \geq S_n(A\bar A)+S_n(B\bar B).
\end{align}
Here $\bar A$ and $\bar B$ are the wedge reflected analogues of the intervals $A$ and $B$ respectively. 
\begin{figure}[!t]
	\hfill
	\subfigure
	{\includegraphics[scale=0.6]{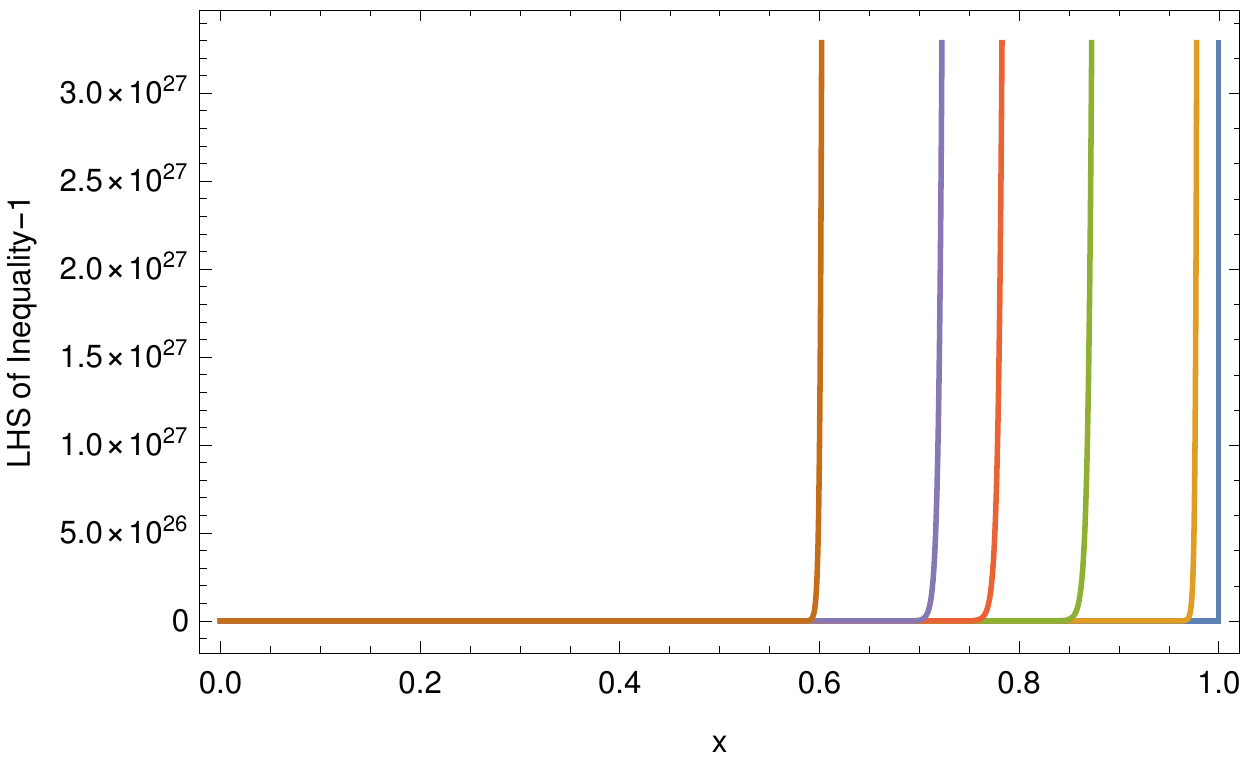}}
	\hfill
	\subfigure
	{\includegraphics[scale=0.57]{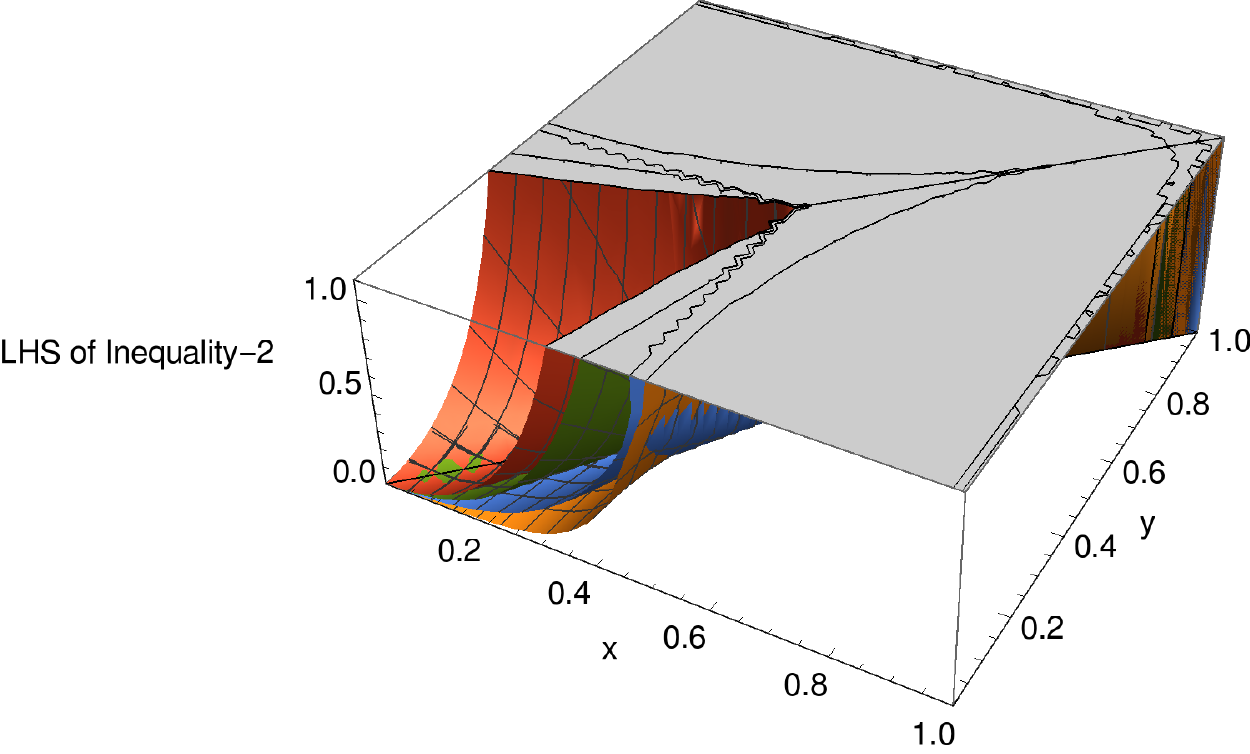}}
	\hfill
	\caption{Verifying the second \Renyi inequalities -- \eqref{ineq1} on left and \eqref{ineq2} on right -- for two intervals, for extremal CFTs with different values of central charges. The first plot is for $k=1,5,10,15,20,45$, (the curves from left to right are for decreasing $k$) while the second one is for $k=1,15,30,45$.}\label{figure-ineq}
\end{figure}

In a chiral CFT, the $n$th \Renyi entropy for two intervals $A=\left[u_1,v_1\right]$ and $B=\left[u_2,v_2\right]$ with cross-ratio $x$, as defined in \eqref{crossrat}, is given by
\begin{align}\label{renyidef}
 S_n= \frac{\kappa^2}{n-1} \left[\frac{c}{12}\left(n-\frac{1}{n}\right)\log\left[x(u_2-v_1)(v_2-u_1)\right]-\log\left[F_n(x)\right]\right],
\end{align}
where $\kappa$ is some constant and $F_n$ is a function of cross-ratio $x$, with $F_n(0)=1$ and $F_n(x)=F_n(1-x)$. Consequently, the $n$-th \Renyi entropies for two intervals obey the following set of inequalities, derivable from \eqref{ineqC} 
\begin{align}
 &\frac{ \partial}{\partial x} \bigg( \frac{ F_n(x) }{(1-x)^{p}} \bigg)  \geq 0 \label{ineq1} \\
 \frac{ F_n(x) }{(1 - x)^p} &\frac{ F_n (y) }{(1-y)^p }-\bigg( \frac{ F_n(z) }{( 1-z)^p }\bigg)^2 \geq 0. \label{ineq2}
\end{align}
Here, $p = c/8$, for the chiral case. In the second inequality, $x$ and $y$ can be anything from 0 to 1, while $z =  {2 \sqrt{x y}}/({1 + \sqrt{1-x} \sqrt{1-y} + \sqrt{xy} })$. The inequalities are not sensitive to the overall constant $\kappa^2$, as they get cancelled out from both sides.


Without the loss of generality we choose, $A=\left[u_1=0,v_1=x\right]  \text{ and } B=\left[u_2=1,v_2=x_{\infty}\right]$. From equation  \eqref{eqs2} we have,
\begin{equation}
 S_2= \kappa^2 \left[\tfrac{c}{8}\log\left[x(1-x)x_{\infty}\right]-\log\left[F_2(x)\right]\right], \text{  where  }  F_2(x)= \frac{1}{\kappa^2}\left[\tfrac{1}{16}x(1-x)\right]^{c/12}Z[\tau(x)] .
\end{equation}
where $\tau(x)$ is given by equation \eqref{taux}. 


The numerical verification of the inequalities \eqref{ineq1} and \eqref{ineq2} however does not yield any surprises. These are satisfied (see Fig.~\ref{figure-ineq}) by the $c=24k$ extremal CFTs. It is worthwhile checking whether the same holds true for the higher \Renyis.

\subsection{On higher genus partition functions}\label{sec:higherGenus}

As mentioned in the introduction, one can use higher genus partition functions for the calculation of \Renyi entropies. The genus-2 partition functions for extremal CFTs have been calculated in \cite{Tuite:1999id,Mason:2006dk,gaiotto-yin,Gaberdiel:2009rd,Gaberdiel:2010jf} and are given in terms of $Sp(4,\mathbb{Z})$ Siegel modular forms. 

The moduli space of a genus $g$ Riemann surface has $6g-6$ real dimensions. There are three presentations in which the genus-2 partition functions can be used. These are (a) the third \Renyi entropy of two intervals, or (b) the second \Renyi entropy of three intervals, or (c) the second \Renyi of a single interval on the torus. The first scenario (a) scans a one-dimensional trajectory in the 6-dimensional moduli space, parametrized by the cross ratio $x$. In (b)  we have a 6-point function of twist operators and this is a 3-dimensional surface in moduli space parametrized by the cross ratios. Finally, in (c) (which is a special case of \S\ref{sec:EEonTheTorus}, albeit non-perturbative) a two-dimensional surface in moduli space is probed -- the moduli being given in terms of the temperature ($\tau=2\pi i\beta/L$) and the interval length ($\ell/L$). The interval length, temperature and the cross-ratios (in the other two cases) can be related to elements of the period matrix of the Siegel modular forms. 

However, translating the results of \cite{Tuite:1999id,Mason:2006dk,gaiotto-yin,Gaberdiel:2009rd,Gaberdiel:2010jf} into \Renyi entropies 
requires a proper handling of conformal anomaly factors\footnote{We are grateful to Ida Zadeh and Xi Yin for discussions on this topic.}; see also \cite{Iqbal:2014tta,Iqbal:2015vka}. One way of calculating these factors is to find the large-$c$ conformal block with fixed internal weights as in \cite{Cho:2017oxl} or from the Liouville action corresponding the relevant correlators as described in \cite{Lunin:2000yv}. It is not clear to us whether closed form results can be obtained (which is a necessity and not an option, for checking the inequalities of \cite{Casini:2010nn}). 


\section{Conclusions and future directions} \label{sec:conclusions}

In this work, we have studied the \Renyi and entanglement entropies of   extremal CFTs. For the case of a single interval on the torus we have considered the $n$th \Renyi entropy in the short interval expansion, verified this with expectations from universal thermal corrections and have also provided a corresponding holographic analysis which is consistent with that of the CFT. The analysis reveals that, for the Monster CFT, features of moonshine are manifested in the \Renyi and entanglement entropies. 
We have also studied the second \Renyi entropy of two intervals on the plane, in which the partition function of the branched Riemann surface can be mapped to the torus partition function.

This is clearly not the full story and we have just begun to scratch the surface (as shown in Fig.~\ref{fig:xkcd}). The computational tractability for the \Renyis~on the torus is rendered by the short interval expansion. Since this approach is perturbative, a better approach to the problem would be welcomed; although a resummation of the series was attempted, but with partial success, in Section \ref{sec:conjecture}. One may expect that further constraints from modular invariance can provide a result for \Renyi and entanglement entropies which is, hopefully, a closed form expression of the modular parameter and the interval length\footnote{We note that it has been shown in \cite{Gaberdiel:2009rd} that higher genus partition functions with $g\leq 4$ for $c\leq 24$ theories are shown to be uniquely fixed by the torus partition function alone.}. 

Another possible approach which we have eschewed in this work, is to evaluate the \Renyi entropy using the Frenkel-Lepowsky-Meurman construction of the moonshine module. This will involve an analysis akin to \cite{Datta:2013hba,Chen:2015cna}; see also section 5 of \cite{Lokhande:2015zma}. The contribution of the bosonic oscillators is fairly easy to handle as it just gets raised to the power of 24. However, dealing with the classical part coming from the compactification on the Leech lattice is non-trivial. A potential drawback of this approach is that the analytic continuation to the $n\to 1$ limit is not known, since the classical part is given in terms of Riemann-Siegel theta functions. Furthermore, it is not known whether the FLM construction has analogues for extremal CFTs with arbitrary $k$. 

There are several other avenues to explore features of entanglement in extremal CFTs. One direction of immediate interest is to evaluate the higher \Renyis~for two intervals on the plane, and verify whether constraints of \cite{Casini:2010nn} are obeyed. This can then shed light on conundrums regarding the existence of extremal CFTs for $k\geq 2$. There are also other bipartitionings of the Hilbert space we can work with. For instance, we can consider a partitioning left v/s right movers  of D-branes having symmetries of $\bM$, which have been constructed in \cite{Craps:2002rw}. It would be intriguing to explore whether the finite piece of the left-right entanglement entropy has a topological interpretation as in \cite{PandoZayas:2014wsa,Das:2015oha,Schnitzer:2015gpa,Zayas:2016drv}. 

It is also worthwhile investigating how the symmetry of the sporadic group $\bM$ arises in the bulk dual. The symmetry is realized as an automorphism of the vertex operator algebra of the CFT. It may be possible to derive an analogue of this symmetry in the gravity dual using the entanglement entropy derived here, along with clues from bulk reconstruction techniques and the proposal for quantum corrections to entanglement \cite{FLM2}. One may also consider an approach based on the bulk-boundary reconstruction of \cite{Hamilton:2006az}.

Finally, it would be exciting to explore entanglement in BPS sectors of string compactifications which exhibit moonshine in their elliptic genus \cite{Eguchi:2010ej,Govindarajan:2009qt,Cheng:2012tq}. Since much remains to be  deciphered regarding moonshine in these setups, it is worthwhile to consider refined measures which can  make these aspects more manifest. The supersymmetric \Renyi entropy \cite{Giveon:2015cgs} (or suitable modifications thereof) can turn out be a useful measure in this context. 

%
%
%
%

\section*{Acknowledgements}
SD thanks Sunil Mukhi, for his seminar at ETH Zurich, and  Roberto Volpato, for conversations which inspired this work. We thank Justin David and Matthias Gaberdiel for discussions and for helpful suggestions on the draft. It is a pleasure to thank Alexandre Belin,  Tarun Grover, Christoph Keller, Wei Li, Roji Pius, Anne Taormina, Roberto Volpato, Edward Witten and Ida  Zadeh for valuable discussions. SD is grateful to  the   organizers and participants of the workshop  \textit{Quantum Gravity and New Moonshines} at the Aspen Center for Physics (supported by NSF grant PHY-160761) for interesting discussions on related topics. SD also thanks the organizers of \textit{Exact methods of low dimensional Statistical Physics}, at the 
Institut d'\'Etudes Scientifiques de Carg\`ese, for an opportunity to present this work.  SP thanks the organizers, participants and lecturers of TASI 2017, especially Miranda Cheng for her  lectures on Moonshine. The work of SD is supported by the NCCR SwissMap, funded by the Swiss National Science Foundation. DD and SP acknowledge the support provided by the US Department of Energy (DOE) under cooperative research agreement DE-SC0009919.

\appendix
\section{Eisenstein series and Ramanujan identities}\label{app:A}

The Eisenstein series are defined by the following lattice sums
\begin{align}
E_{2\k}(\tau ) = \sum_{(m,n)\in \Lambda \backslash (0,0) } \frac{1}{(m+n\tau)^{2\k}}
\end{align}
These are weight $2\k$ modular forms (for $\k\geq 2$). $E_2$ is mock modular. Their $q$-series are given by 
\begin{align}
E_{2\k} (\tau) = 1- \frac{4\k}{B_{2\k}} \sum_{n=1}^{\infty} \sigma_{2\k-1}(n) q^n .
\end{align}
Here, $B_r$ is the Bernoulli number and $\sigma_{p}(s)$ is the divisor function. The Eisenstein series fall under the category of holomorphic forms \ie these are modular forms for which non-negative powers of $q$ in the $q$-series do not exist. $E_4$ and $E_6$, in particular, are algebraically independent and generate the space of all modular forms.

The Klein $j$-invariant is a weakly holomorphic modular form of weight zero (\ie it has finitely many negative powers of $q$ in the $q$-series). It can be written in terms of the Eisenstein series.  
\begin{align}\label{j-in-E}
 j(\tau) =  \frac{1728E_4^3 }{ E_4^3 - E_6^2 } 
\end{align}
We use the standard convention for the nome,  $q=e^{2\pi i \tau}$.

The cusp forms are modular forms having solely positive powers of $q$ in the $q$-series. The lowest weight cusp form is the discriminant (or the Ramanujan tau function). This is given by
\begin{align}
\Delta(\tau) = \eta(\tau)^{24} = {E_4^3 - E_6^2 \over 1728}, 
\end{align}
and has weight 12. The lowest weight is 12 since it is the first instance by which the $q^0$ term can be killed by a combination of powers of $E_4$ and $E_6$ (\ie the lowest weight is the lowest common multiple of the weights of generators $E_{4,6}$). 

The Ramanujan identities provide expressions for the $\tau$-derivatives of the Eisenstein series. These are 
\begin{align}\label{Ramanujan}
\begin{aligned}
q\, \partial_q \,  E_2 \, = \, \frac{E_2^2-E_4}{12}, \qquad
q\, \partial_q \,  E_4 \, = \, \frac{E_2 E_4-E_6}{3}, \qquad 
q\, \partial_q \,  E_6 \, = \, \frac{E_2 E_6-E_4^2}{2}. 
\end{aligned}
\end{align}
These relations have been used in the final expressions for the expectation values of quasi-primaries on the torus. This and \eqref{j-in-E} also enables one to write a simple expression for the derivative of the $j$-invariant
\begin{align}\label{j-magic}
q \, \partial_q \, j \, = \, -\, \frac{E_6}{E_4}\,  j . 
\end{align}
This leads to simplifications in the expressions of the 1-point functions of quasiprimaries \eg \eqref{st-k1} and \eqref{monster-1-point}.

\section{Extremal CFT at $c=48$ and $c=72$}\label{app:4872}
\subsection*{Extremal CFT at $c=48$}

The $k=2$ (i.e. $c=48$) extremal CFT on a torus is described by the partition function
\begin{align}
\nonumber Z_{k=2}(\tau) &= j(\tau )^2-1488 j(\tau )+159769,\\
&=\frac{1}{q^2}+1+42987520 q+40491909396 q^2+O\left(q^3\right)
\end{align}
The $q$ expansion of the partition function reveals that the lowest non-trivial quasi-primary is $T$, having conformal weight $2$. There are $42987520$ operators of conformal weight $3$, out of which, $42987519$ are primary operators and the lone descendant is $\partial T$. Hence, the leading contribution to \Renyi entropy $S_n$ at low temperature comes from $T$ alone
\begin{align}\label{s-tx}
\delta S_n=\delta S_n ^{(T)} = - \frac{c(n+1)^2(n-1)}{18n^3} \sin^4 (\tfrac{\pi \ell}{L}) q^2 +  \frac{1}{1-n} \bigg[\frac{1}{n^{3}} \frac{\sin^{4} (\pi \ell /L)}{\sin^{4} (\pi \ell /nL)} -n \bigg]q^2
\end{align}

One can match the small $\ell$ limit of the equation  \eqref{s-tx} with the one coming from the short interval expansion. As elucidated earlier for the Monster CFT, the short interval expansion provides us with an expression, perturbative in $\ell$, but non-perturbative at each order in the modular parameter $\tau$ of the torus. Subsequently, the matching is done upon doing a small $q$ expansion of the result. This involves (as in the case for $k=1$) evaluation of the torus one point function of quasi-primaries in the identity module. 

For the $k=2$ extremal CFT, the expectation values appearing in the short interval expansion, are given by:
\begin{align} \label{expqp}
\vev{T}&=\frac{8 \pi ^2E_6}{E_4} \frac{j(j-744) }{ j^2-1488 j+159769} \\
\vev{\cA}&=\frac{8 \pi ^4}{15} \left(\frac{E_4 (31j^2-23808j+159769)}{(j^2-1488j+159769)}+\frac{20 E_6^2 (5 j^2-1488j)}{ E_4^2 (j^2-1488j+159769)}\right)\\
\vev{\cB}&=-\frac{32 \pi ^6E_6}{175} \left(\frac{31j^2-30504j+1597690}{ j^2-1488j+159769}\right)\\
\vev{\cD}&=\frac{16 \pi ^6E_6}{30501} \left(\frac{714705 j^2-259500504j+15657362}{j^2-1488j+159769}+\frac{108448 E_6^2(5 j^2-372j)}{E_4^3 (j^2-1488j+159769)}\right)
\end{align}

Using the short interval expansion as described in Sec \ref{sec:SIE}, the expectation values of quasi-primaries lead to an expression for \Renyi entropy
\begin{align}
S_n = \frac{4(n+1)}{n} \log \frac{\ell}{\epsilon} + \sum_{\k=1}^\infty \, {\cX_{2\k}(\tau) \over Z_2(\tau)^\k}  \left(\ell \over L\right)^{2\k}.
\end{align}
Here, $\cX_{2\k}(\tau)$ are the meromorphic weakly modular functions of weight $2\k$ and given by
\begin{align}
\cX_{2}(\tau)&=-\frac{2 \pi ^2}{3}\frac{n+1}{n} \frac{E_6 }{E_4}\mathscr{P}_{2,1}\\
\cX_{4}(\tau)&=\frac{\pi ^4}{1080}\frac{n+1}{n^3}\left[\frac{E_6^2}{E_4^2}\mathscr{P}_{4,1}-E_4 Z_{2}\mathscr{P}_{4,2}\right]\\
\cX_{6}(\tau) &=\frac{\pi ^6}{102060}\frac{n+1}{n^5}\left[\frac{E_6^3}{E_4^3}\mathscr{P}_{6,1}-E_6 Z_2\mathscr{P}_{6,2}\right]
\end{align}
where we have defined the following polynomials of the $j$-invariant
\begin{align}
\mathscr{P}_{2,1} = \sum_{k=0}^{ 2}f^{2,1}_{k}j^{k+1}, \quad   \text{and }\ 
 \mathscr{P}_{m,1} =  \sum_{k=0}^{m-1}f^{m,1}_k j^{k+1} , \  \mathscr{P}_{m,2}=\sum_{k=0}^{m-2} f^{m,2}_kj^{k}  \quad \text{for } m=4,6. \nn
\end{align}
%
%
The coefficients of the polynomials above are the following:
\begin{footnotesize}
\begin{align*}
f^{2,1}_2&=1, \qquad f^{2,1}_1=744\\
f^{4,1}_3&=2(19  n^2-31),\qquad f^{4,1}_2=2(1488 n^2+16368 )\\
f^{4,1}_1&=2(5611004  n^2-12253436), \qquad f^{4,1}_{0}=4754725440 (n^2-1)\\
f^{4,2}_2&=62(n^2-1),\qquad f^{4,2}_1=-47616(n^2-1)\\
f^{4,2}_0&=319538(n^2-1)\\
f^{6,1}_5&=-(237 n^4-1023n^2+930), \qquad f^{6,1}_4=-(866760 n^4-1258104 n^2+69936)\\
f^{6,1}_3&=-(106931498 n^4-443936764 n^2+576132818)\\
f^{6,1}_2&=-(237752546256 n^4-461214060768  n^2+164157881616)\\
f^{6,1}_1&=(267774584843024  n^4-624694566015520  n^2+356919981172496)\\
f^{6,1}_0&=5317604101763520 (n^2-1)^2.
\end{align*}
\end{footnotesize}
We define $f^{6,2}_j \equiv (n^2-1)\tilde{f}^{6,2}_j$, where $\tilde{f}^{6,2}_{j}$s are given by 
\begin{footnotesize}
\begin{align*}
\tilde{f}^{6,2}_4&=-93(n^2-10),\qquad \tilde{f}^{6,2}_3=-(1165848n^2+153264)\\
\tilde{f}^{6,2}_2&=(465242371 n^2+78648104), \qquad \tilde{f}^{6,2}_1=-(212892831576  n^2-196845633216)\\
\tilde{f}^{6,2}_0&=(484996533859 n^2-25526133361).
\end{align*}
\end{footnotesize}

\subsection*{Extremal CFT at $c=72$}
The partition function for the $k=3$ (\ie  $c=72$) extremal CFT on a torus is given by
\begin{align}
\nonumber Z_{k=3}(\tau) &= j(\tau)^3-2232 j(\tau)^2+1069957 j(\tau)-36867719,\\
&=1/q^3 + 1/q + 1 + 2593096794 q + 12756091394048 q^2 + O(q^3)
\end{align}
From the $q$ expansion, it is evident that the lightest non-trivial quasi-primary is $T$, having conformal weight $2$. The only operator, appearing with conformal weight $3$, is $\partial T$. The extremal CFTs have lightest primaries appearing with conformal weight $k+1$ ($4$ in this case), by construction. The number of operators with conformal weight $4$ is $2593096794$, out of which, $2593096792$ ones are the primaries while the rest of the two are $\cA$ and $\partial^2T$. Here, as well, the leading contribution to \Renyi entropy $S_n$ at low temperature comes from the stress tensor.
\begin{align}\label{s-t1}
\delta S_n=\delta S_n ^{(T)} = - \frac{c(n+1)^2(n-1)}{18n^3} \sin^4 (\tfrac{\pi \ell}{L}) q^2 +  \frac{1}{1-n} \bigg[\frac{1}{n^{3}} \frac{\sin^{4} (\pi \ell /L)}{\sin^{4} (\pi \ell /nL)} -n \bigg]q^2
\end{align}

In fact, this is a generic feature of all $k\geq 2$ extremal CFTs, the leading contribution comes from $T$ alone. We will not repeat the matching here, since this has been shown generally in Sec \ref{sec:otherExtremal}. Nonetheless for the sake of completeness, we provide the torus one point functions of quasi-primaries, appearing in short interval expansion, for the $k=3$ case; these are given by
\begin{align}
\vev{T}&=\tfrac{4 \pi ^2 E_6}{E_4}\left(\tfrac{3 j^3-4464 j^2+1069957j}{j^3-2232 j^2+1069957 j-36867719}\right)\\
\vev{\cA}&=\tfrac{4 \pi ^4}{15}\left(\tfrac{3 E_4(31 j^3-46872 j^2+11769527 j-36867719)}{\left(j^3-2232 j^2+1069957 j-36867719\right)}+\tfrac{40E_6^2  (12 j^3-11160 j^2+1069957j)}{E_4^2 (j^3-2232 j^2+1069957 j-36867719)}\right)\\
\vev{\cB}&=\tfrac{-48 \pi ^6E_6}{175}\left(\tfrac{31 j^3-53568 j^2+18189269 j-368677190}{j^3-2232 j^2+1069957 j-36867719}\right)\\
\vev{\cD}&=\tfrac{8 \pi ^6 E_6}{45621}\left(\tfrac{4904541 j^3-4854564288 j^2+578130935767 j-21457012458}{j^3-2232 j^2+1069957 j-36867719}+\tfrac{81104 E_6^2 (84 j^3-44640 j^2+1069957j)}{E_4^3(j^3-2232 j^2+1069957 j-36867719)}\right)
\end{align}

The \Renyi entropy reads as follows:
\begin{align}
S_n = \frac{6(n+1)}{n} \log \frac{\ell}{\epsilon} + \sum_{\k=1}^\infty \, {\cX_{2\k}(\tau) \over Z_3(\tau)^\k}  \left(\ell \over L\right)^{2\k} 
\end{align}
where the weak modular function of weight $2k$ appears again and given by the following expressions for $k=3$:
\begin{align}
\cX_{2}(\tau)&=-\frac{\pi ^2}{3}\frac{n+1}{n} \frac{E_6}{E_4}\mathscr{P}_{2,1}\\
\cX_{4}(\tau)&=\frac{\pi ^4}{3240}\frac{n+1}{n^3} \left[\frac{E_6^2}{E_4^2}\mathscr{P}_{4,1}-E_4Z_3\mathscr{P}_{4,2}\right]\\
\cX_{6}(\tau) &=\frac{\pi ^6}{1837080 } \frac{n+1}{n^5}\left[\frac{E_6^3}{E_4^3 }\mathscr{P}_{6,1}-E_6Z_3\mathscr{P}_{6,2}\right]
\end{align}
where the polynomials of $j$ are defined as 
\begin{align}
\mathscr{P}_{2,1}&=j\left(\sum_{k=0}^{k=2}f^{2,1}_{k}j^{k}\right);\\
 \mathscr{P}_{4,1}&=j\left(\sum_{k=0}^{k=5}f^{4,1}_k j^k\right), \qquad  \mathscr{P}_{4,2}=\left(\sum_{k=0}^{k=3} f^{4,2}_kj^{k}\right); \\
\mathscr{P}_{6,1}&=j\left(\sum_{k=0}^{k=8}f^{6,1}_{k}j^{k}\right), \qquad \mathscr{P}_{6,2}=\left(\sum_{k=0}^{k=6}f^{6,2}_{k}j^{k}\right).
\end{align}
Here $f^{i,j}_{k}$s are function of the \Renyi index $n$ only and are given by: 
\makeitsmall
\begin{align*}
f^{2,1}_2&=3, \qquad f^{2,1}_{1}=-4464, \qquad f^{2,1}_0=1069957\\
f^{4,1}_5&=(171 n^2-279), \qquad f^{4,1}_4=-(241056n^2-562464)\\
 f^{4,1}_3&=(57892482 n^2-374056938), \qquad f^{4,1}_2=(62642091456n^2+51988821696)\\
  f^{4,1}_1&=(18170421644291n^2-31908117426479), \qquad f^{4,1}_0=4733624882169960(n^2-1)\\
f^{4,2}_3&=279(n^2-1), \qquad f^{4,2}_2=-421848(n^2-1)\\
f^{4,2}_1&=105925743(n^2-1), \qquad f^{4,2}_0=-331809471(n^2-1).
\end{align*}
\begin{align*}
f^{6,1}_8&=-(6399 n^4-27621 n^2+25110), \qquad f^{6,1}_7=-(9119952 n^4+27681264 n^2-54157248)\\
f^{6,1}_6&=(42500558241 n^4-34278659631 n^2-38207667042)\\
f^{6,1}_5&=-(49955320651536 n^4-79354333120896 n^2+4209289143408) \\
f^{6,1}_4&=(16917854014928979 n^4-32734292306379489 n^2+5121944013384702)\\
f^{6,1}_3&=(6046225178420247156 n^4-13165871496849172584 n^2+9327348981409665780)\\
f^{6,1}_2&=(1285155631058486117707 n^4-2526894768482940564821 n^2+1065354212232184136122)\\
f^{6,1}_1&=-(580982466026811352729428 n^4-1289597264020531187530200  n^2\\
&\ \ \ +708614797993719834800772)\\
f^{6,1}_0&=3664876992152254105946040 (n^2-1)^2.
\end{align*}
\endmakeitsmall
We define $f^{6,2}_{j} \equiv (n^2-1)\tilde{f}^{6,2}_j$, where $\tilde{f}$ is given by
\makeitsmall
\begin{align*}
\tilde{f}^{6,2}_6&=-2511(n^2-10), \qquad \tilde{f}^{6,2}_5=-9(3082392 n^2+4611312)\\
\tilde{f}^{6,2}_4&=9(6555493484 n^2+1365872932), \qquad \tilde{f}^{6,2}_3=-9(4344636377193 n^2-1217158713330)\\
\tilde{f}^{6,2}_2&=9(491636422030439 n^2-49946709507086)\\
\tilde{f}^{6,2}_1&=-9(72148332579073807n^2-65402917121981614)\\
\tilde{f}^{6,2}_0&=9(77476036142988777 n^2-4077686112788883)
\end{align*}
\endmakeitsmall
%

{
\hypersetup{urlcolor= RoyalBlue!20!black}
\setlength{\parskip}{2pt}

\bibliographystyle{bibstyle2017}
\bibliography{collection}

\hypersetup{urlcolor=RoyalBlue!60!black}
}


\end{document}